\documentclass[aps,prd,preprintnumbers,superscriptaddress,showpacs,showkeys,floatfix,twocolumn,nofootinbib]{revtex4-2}

\usepackage{amsmath,amssymb,amsfonts,dcolumn}
\usepackage{graphicx}

\usepackage[dvipsnames]{xcolor}
\usepackage{hyperref}

\usepackage{fix-cm}

\usepackage{slashed} % Dirac Slash Notation

\usepackage{booktabs,tabulary}
\usepackage{multirow}

\newcommand{\beq}{\begin{equation}}
\newcommand{\eeq}{\end{equation}}

\newcommand{\LN}[1]{\textsc{#1}} % Last names in small caps
\newcommand{\Lat}[1]{\textit{#1}} % Latin phrases in italic (e.g., i.e. etc.)

\newcommand{\MetaPrime}{M_{\eta'}}
\newcommand{\GetaPrime}{\Gamma_{\eta'}}
\newcommand{\Meta}{M_\eta}
\newcommand{\MpiN}{M_{\pi^0}}
\newcommand{\MpiC}{M_{\pi^\pm}}

\newcommand{\MK}{M_K}
\newcommand{\MKStar}{{M_{K^*}}}

\newcommand{\Mrho}{M_\rho}
\newcommand{\MrhoPrime}{M_{\rho'}}
\newcommand{\Grho}{\Gamma_\rho}
\newcommand{\GrhoPrime}{\Gamma_{\rho'}}

\newcommand{\Momega}{M_\omega}
\newcommand{\MomegaPrime}{M_{\omega'}}
\newcommand{\Gomega}{\Gamma_\omega}
\newcommand{\GomegaPrime}{\Gamma_{\omega'}}

\newcommand{\Mphi}{M_\phi}
\newcommand{\MphiPrime}{M_{\phi'}}
\newcommand{\Gphi}{\Gamma_\phi}
\newcommand{\GphiPrime}{\Gamma_{\phi'}}

\newcommand{\massIn}{M_{\eta^{(\prime)}}}
\newcommand{\massOut}{M_{\pi^0/\eta}}
\newcommand{\GammaIn}{\Gamma_{\eta^{(\prime)}}}

\newcommand{\M}{\mathcal{M}}
\newcommand{\Hel}{H}
\newcommand{\F}{\mathcal{F}}
\newcommand{\Q}{\mathcal{Q}}

\newcommand{\BR}{\mathcal{B}}

\newcommand{\had}{\text{H}}
\newcommand{\QED}{\text{QED}}
\newcommand{\resc}{\text{resc}}
\newcommand{\Born}{\text{Born}}

\newcommand{\PP}{\mathrm{P}}
\newcommand{\CC}{\mathrm{C}}
\newcommand{\TT}{\mathrm{T}}
\newcommand{\GG}{\mathrm{G}}

\providecommand{\keV}{\,\text{keV}}
\providecommand{\MeV}{\,\text{MeV}}
\providecommand{\GeV}{\,\text{GeV}}

\newcommand{\perc}{\%}

\renewcommand{\Im}{\text{Im}\,}
\renewcommand{\Re}{\text{Re}\,}
\newcommand{\Tr}{\text{Tr}\,}
\newcommand{\diag}{\text{diag}\,}

\newcommand{\iu}{\mathrm{i}}
\newcommand{\dfk}{\mathrm{d}^4k\,}
\newcommand{\dff}{\mathrm{d}}

\newcommand{\Uthree}{\mathrm{U}(3)}

\newcommand{\Order}{\mathcal{O}} 

\newcommand{\sgn}{\mathrm{sgn}}

\renewcommand{\max}{\text{max}}
\newcommand{\thr}{\text{thr}}

\newcommand{\eps}{\epsilon}

\newcommand{\BW}{\text{BW}}
\newcommand{\disp}{\text{disp}}

\begin{document}
%---------------------------------------------------------------------------------------------------

%These lines have to go after \begin{document} and adjust the referenced labels for \autoref.
%---------------------------------------------------------------------------------------------------
\renewcommand{\figureautorefname}{Fig.}
\renewcommand{\tableautorefname}{Table}
\renewcommand{\chapterautorefname}{Ch.}
\renewcommand{\sectionautorefname}{Sec.}
\renewcommand{\subsectionautorefname}{Sec.}
\renewcommand{\appendixautorefname}{App.}
\def\equationautorefname~#1\null{Eq.~(#1)\null}
%---------------------------------------------------------------------------------------------------

\title{The semileptonic decays \boldmath{$\eta^{(\prime)} \to \pi^0 \ell^+ \ell^-$} and \boldmath{$\eta' \to \eta \ell^+ \ell^-$} in the standard model}

\author{Hannah Schäfer}
\email{schaefer@hiskp.uni-bonn.de}

\author{Marvin Zanke}
\email{zanke@hiskp.uni-bonn.de}

\author{Yannis Korte}
\email{korte@hiskp.uni-bonn.de}

\author{Bastian Kubis}
\email{kubis@hiskp.uni-bonn.de}

\affiliation{Helmholtz-Institut für Strahlen- und Kernphysik (Theorie) and\\ Bethe Center for Theoretical Physics, Universität Bonn, 53115 Bonn, Germany}

\begin{abstract}
We perform a theoretical analy\-sis of the semileptonic decays $\eta^{(\prime)} \to \pi^0 \ell^+ \ell^-$ and $\eta' \to \eta \ell^+ \ell^-$, where $\ell = e, \mu$, via a charge-conjugation-conserving two-photon mechanism.
The underlying form factors are modeled using vector-meson dominance, phenomenological input, and $\Uthree$ flavor symmetry. 
We consider both a monopole and a dipole model, the latter tailored such that the expected high-energy behavior is ensured.
Furthermore, we benchmark the effect of $S$-wave rescattering contributions to the decays.
We infer significant effects of the form factors neglected in the literature so far, still finding branching ratios of the various decays well below the current experimental upper limits.
\end{abstract}


\maketitle

%---------------------------------------------------------------------------------------------------
\section{Introduction}
\label{sec:intro}
%---------------------------------------------------------------------------------------------------
Within the standard model (SM) of particle physics, the strong and electromagnetic interactions conserve the symmetries parity ($\PP$), charge conjugation ($\CC$), and time reversal ($\TT$) separately.
For this reason, the decays $\eta^{(\prime)} \to \pi^0 \ell^+ \ell^-$ and $\eta' \to \eta \ell^+ \ell^-$ can---mediated via the strong and electromagnetic force---only proceed via a $\CC$-even two-photon mechanism due to $\CC(\eta^{(\prime)}) = \CC(\pi^0) = +1$; \Lat{i.e.}, they appear as one-loop processes at lowest order.\footnote{%
    Contributions from the weak interactions are also required to vanish at tree level.}
As a result, the SM contribution to those decays is strongly suppressed, rendering them well-suited candidates for searches for physics beyond the SM (BSM).
In fact, BSM contributions to the discussed decays, either mediated via a $\CC$-odd one-photon exchange~\cite{Bernstein:1965hj,Barrett:1965ia,Akdag:2022sbn,Akdag:2023pwx} or due to other BSM mechanisms such as new light scalars~\cite{Gan:2020aco} and unconventional sources of $\CC\PP$~violation~\cite{Escribano:2022zgm}, are themselves subject to ongoing analyses.

Historically, calculations of $\eta \to \pi^0 \ell^+ \ell^-$ were based on different models for the $\eta \to \pi^0 \gamma^* \gamma^*$ vertex function, as the conversion $\gamma^* \gamma^* \to \ell^+ \ell^-$ depends solely on quantum electrodynamics (QED) and is, hence, in principle straightforward.
This is not unlike the rare dilepton decays of the lightest flavor-neutral pseudoscalars, $P \to \ell^+ \ell^-$, $P = \pi^0$, $\eta$, $\eta'$, similarly loop-induced and completely calculable once the corresponding transition form factors $P \to \gamma^* \gamma^*$ are known; see Refs.~\cite{Masjuan:2015cjl,Weil:2017knt,Hoferichter:2021lct} for recent work and references therein.
For these decays, a reasonable behavior of the transition form factors for large photon virtualities is not only a requirement for a precision calculation, but a necessity to regularize the otherwise ultraviolet-divergent loop integral.
This was similarly realized in early theoretical work on $\eta \to \pi^0 \ell^+ \ell^-$ in the late 1960s, which was based on the simplest possible, point-like effective operator for $\eta \to \pi^0 \gamma \gamma$~\cite{LlewellynSmith:1967,Smith:1968}: the loop was rendered finite either with an \Lat{ad-hoc} form factor~\cite{LlewellynSmith:1967} or reconstructed dispersively from the unambiguously calculable imaginary part, using a finite energy cutoff~\cite{Smith:1968}.
As the effective operator only contained $S$-wave interactions in either case---leading to helicity suppression of the resulting dilepton mechanism---these calculations only determined a subdominant contribution, underestimating in particular the $\eta \to \pi^0 e^+ e^-$ rate by orders of magnitude.
\begin{table*}[t]
\nocite{Cheng:1967zza,Ng:1992yg,Ng:1993sc,ParticleDataGroup:2022pth}
	\centering
	\begin{tabular}{l  r  c  r}
	\toprule
		& Branching ratio & Ancillary information & Reference \\
		\midrule
  		$\eta \to \pi^0 e^+ e^-$ & $9.9 \times 10^{-9}$ & VMD model & \cite{Cheng:1967zza} \\
        $\eta \to \pi^0 e^+ e^-$ & $8.4^{+4.6}_{-3.8} \times 10^{-10}$ & Unitarity bounds, VMD model & \cite{Ng:1992yg} \\
        $\eta \to \pi^0 e^+ e^-$ & $9.2(1.5) \times 10^{-10}$ & Quark-box model, $m_q = 330 \MeV$ & \cite{Ng:1993sc} \\
        \multirow{2}{*}{$\eta \to \pi^0 \mu^+ \mu^-$} & $3.8^{+2.3}_{-1.5} \times 10^{-10}$ & Unitarity bounds, VMD model & \multirow{2}{*}{\cite{Ng:1992yg}} \\
        & $6.9^{+4.6}_{-3.8}  \times 10^{-10}$ & As above, supplemented by $a_0$ & \\
        $\eta \to \pi^0 \mu^+ \mu^-$ & $3.3(5) \times 10^{-9}$ & Quark-box model, $m_q = 330 \MeV$ & \cite{Ng:1993sc} \\
        \midrule
        $\eta \to \pi^0 e^+ e^-$ & $< 7.5 \times 10^{-6}$ & $3 \times 10^7$ $\eta$ events & WASA-at-COSY~\cite{WASA-at-COSY:2018jdv} \\
        $\eta \to \pi^0 \mu^+ \mu^-$ & $< 5 \times 10^{-6}$ & $2 \times 10^7$ $\eta$ events & Dzhelyadin \Lat{et al.}~\cite{Dzhelyadin:1980ti} \\
        $\eta' \to \pi^0 e^+ e^-$ & $< 1.4 \times 10^{-3}$ & $1.3 \times 10^6$ $\eta'$ events & CLEO~\cite{CLEO:1999nsy} \\
        $\eta' \to \pi^0 \mu^+ \mu^-$ & $< 6 \times 10^{-5}$ & $10^7$ $\eta'$ events & Dzhelyadin \Lat{et al.}~\cite{Dzhelyadin:1980ti} \\
        $\eta' \to \eta e^+ e^-$ & $< 2.4 \times 10^{-3}$ &$1.3 \times 10^6$ $\eta'$ events & CLEO~\cite{CLEO:1999nsy} \\
        $\eta' \to \eta \mu^+ \mu^-$ & $< 1.5 \times 10^{-5}$ & $10^7$ $\eta'$ events & Dzhelyadin \Lat{et al.}~\cite{Dzhelyadin:1980ti} \\
		\bottomrule
	\end{tabular}
	\caption{Historical theoretical results on the branching ratios for $\eta \to \pi^0 \ell^+ \ell^-$ and experimental upper limits for the different decay channels $\eta^{(\prime)} \to [\pi^0/\eta] \ell^+ \ell^-$, the latter all at $90\perc$ confidence level.
    Note that, for reasons of consistency with the experimental upper limits, we converted the theoretical results from decay widths to branching ratios by using an up-to-date central value~\cite{ParticleDataGroup:2022pth} for the $\eta$ width; see also \autoref{tab:constants}.}
	\label{tab:historical_results}
\end{table*}

\begin{figure}[t]
	\centering
	\includegraphics[scale=0.9]{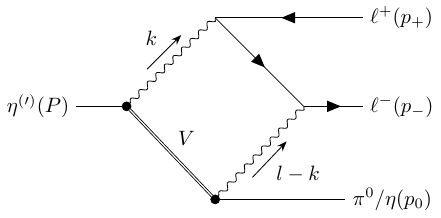}
    \\[1cm]
	\includegraphics[scale=0.9]{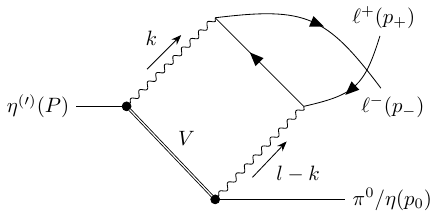}
	\caption{The $t$- (\textit{top}) and $u$-channel (\textit{bottom}) diagrams that contribute to $\eta^{(\prime)} \to [\pi^0/\eta] \ell^+ \ell^-$ under the assumption that the underlying two-photon amplitudes are dominated by the exchange of the vector mesons $V = \rho, \omega, \phi$.}
	\label{fig:fd_process}
\end{figure}
On the other hand, a first vector-meson-dominance (VMD) model calculation~\cite{Cheng:1967zza}, which based the $\eta \to \pi^0 \gamma \gamma$ amplitude on $\rho$ and $\omega$ exchange, $\rho \equiv \rho^0(770)$, $\omega \equiv \omega(782)$, required no such further regularization: the additional vector-meson propagators, singularities in the crossed channels providing so-called left-hand cuts, dampen the high-energy behavior sufficiently such that the loop integral is convergent; see \autoref{fig:fd_process}.
The coupling constants for the $V \to P \gamma$ transitions, $V = \rho, \omega$, $P = \eta, \pi^0$, largely unknown at the time, had to be estimated in a quark model.
In this way, realistic rates $\BR(\eta \to \pi^0 e^+ e^-)/\BR(\eta \to \pi^0 \gamma \gamma) \approx 10^{-5}$ were obtained.
In the 1990s, the two decays $\eta \to \pi^0 e^+ e^-$ and $\eta \to \pi^0 \mu^+ \mu^-$ were reconsidered by calculating unitarity bounds~\cite{Ng:1992yg,Ng:1993sc}.
These are based on the observation that the amplitude $\eta \to \pi^0 \gamma \gamma$ (with real photons) model-independently determines the imaginary part of the dilepton amplitudes, thus providing a lower limit to the corresponding rates.
The diphoton decays were calculated in VMD, supplemented with scalar $a_0(980)$ exchange~\cite{Ng:1992yg}, or based on a constituent-quark-box model~\cite{Ng:1993sc}.
The numerical results of these older calculations are collected in \autoref{tab:historical_results}.

Today, we understand the mechanism for $\eta \to \pi^0 \gamma \gamma$ (and the related $\eta'$ decays) much better, while precision calculations are still a challenge.
Chiral perturbation theory~\cite{Gasser:1984gg} allows us to understand this reaction in terms of a systematic expansion at low momenta: the dominant contribution originates from a set of next-to-next-to-leading-order counterterms~\cite{Ametller:1991dp,Jetter:1995js}, whose size can phenomenologically be estimated in terms of vector-meson exchanges.
The resulting predictions agree with the data~\cite{Prakhov:2005vx,Prakhov:2008zz,A2atMAMI:2014zdf} rather well~\cite{Danilkin:2017lyn}, and rescattering corrections in the scalar channel~\cite{Oset:2002sh,Oset:2008hp} are moderate in size~\cite{Lu:2020qeo}.
Similarly, vector-meson exchanges dominate the decays $\eta' \to \pi^0 \gamma \gamma$ and $\eta' \to \eta \gamma \gamma$~\cite{Escribano:2018cwg}, with only minor $S$-wave corrections to the $\gamma \gamma$ spectra.

The most recent theoretical work on the decays $\eta^{(\prime)} \to \pi^0 \ell^+ \ell^-$ and $\eta' \to \eta \ell^+ \ell^-$~\cite{Escribano:2020rfs} employs this modern knowledge to a large extent.
It once more models the two-photon amplitudes with a VMD ansatz, superseding Ref.~\cite{Cheng:1967zza} by retaining all lepton mass effects and Ref.~\cite{Ng:1992yg} by calculating the real parts of the amplitudes explicitly; the current phenomenological information on vector--pseudoscalar--photon couplings is used therein.
Perhaps surprisingly, what has still not been implemented is the dependence on the photon virtualities, \Lat{i.e.}, the vector-to-pseudoscalar transition form factors~\cite{Landsberg:1985gaz,Fang:2021wes}.
These have garnered significant interest in the last few years, both phenomenologically~\cite{Terschlusen:2010gtc,Terschlusen:2012xw,Schneider:2012ez,Danilkin:2014cra} and, in particular for the $\rho \to \pi$ transition form factor, on the lattice~\cite{Briceno:2015dca,Briceno:2016kkp,Alexandrou:2018jbt,Niehus:2021iin}.
Furthermore, the behavior of these form factors for asymptotically large momentum transfers is known~\cite{Farrar:1975yb,Vainshtein:1977db,Lepage:1979zb,Lepage:1980fj,Chernyak:1983ej}.
This is the major novelty of this article and the main advance compared to Ref.~\cite{Escribano:2020rfs}: by providing a realistic model for $\eta^{(\prime)} \to \pi^0 \gamma^* \gamma^*$ and $\eta' \to \eta \gamma^*\gamma^*$, including the dependence on the photon virtualities, we are able to give a more reliable prediction for the rates of the corresponding dilepton decays in the SM.
Furthermore, by lifting the (somewhat artificial) dependence of the loop regularization on the left-hand cuts, we can, for the first time, also test the effect of $S$-wave rescattering contributions.
Varying the form-factor models allows us to assess the remaining theoretical uncertainties of our predictions.

Experimentally, the decay $\eta \to \pi^0 e^+ e^-$ has been searched for since the 1960s~\cite{Rittenberg:1965zz, Bazin:1968zz, Jane:1975nt}, motivated by the search for possible $\CC$~violation in the strong and electromagnetic interactions.
To date, only upper limits have been established for all decays studied in this article, the most rigorous ones being collected in Table~\ref{tab:historical_results}.\footnote{%
    Note that those upper limits were obtained assuming a flat \LN{Dalitz}-plot distribution, which our results indicate to be an insufficient assumption; see the discussion in \autoref{sec:discussion_diff} below.}
The most stringent upper limits, those for $\eta \to \pi^0 e^+ e^-$ from WASA-at-COSY~\cite{WASA-at-COSY:2018jdv} and for $\eta \to \pi^0 \mu^+ \mu^-$ from Lepton-G~\cite{Dzhelyadin:1980ti}, are still more than three orders of magnitude above the theoretical SM branching ratios; for the $\eta'$ decays, this margin is even larger.
There is, even so, the prospect of improved experimental results by the REDTOP collaboration~\cite{Gatto:2019dhj, REDTOP:2022slw}, which plans to search for rare decays with an unprecedented number of $\eta$ and $\eta'$ events.

This article is structured as follows.
In \autoref{sec:amplitudes}, we construct the amplitudes for the decays $\eta^{(\prime)} \to \pi^0 \ell^+ \ell^-$ and $\eta' \to \eta \ell^+ \ell^-$ as well as the corresponding two-photon analogs, with the latter serving as normalization channels.
For the semileptonic decays, a set of form factors that incorporate the non-perturbative physics of the process is introduced and their normalizations are determined from phenomenological input. 
These form factors are then parameterized in \autoref{sec:form_factors} by means of two distinct VMD models, including the construction of dispersively improved variants.
In \autoref{sec:observables}, we discuss the calculation of observables---branching ratios as well as differential distributions---via a \LN{Passarino}--\LN{Veltman} (PV) decomposition.
Scalar rescattering contributions are analyzed in \autoref{sec:rescattering}.
Our numerical results are discussed in \autoref{sec:results_discussion}, and we summarize our findings in \autoref{sec:summary}.
Further details are provided in the appendices.

%---------------------------------------------------------------------------------------------------
\section{Amplitudes}
\label{sec:amplitudes}
%---------------------------------------------------------------------------------------------------
\begin{table}[t]
	\centering
	\begin{tabular}{l  r  r}
	\toprule
		& $\Gamma$/\keV~\cite{ParticleDataGroup:2022pth} & $\lvert \F_{V P}(0) \rvert / \GeV^{-1}$ \\
		\midrule
        $\rho \to \pi^0 \gamma$ & $69(12)$ & $0.73(6)$ \\
		$\omega \to \pi^0 \gamma$ & $725(26)$ & $2.33(4)$ \\
		$\phi \to \pi^0 \gamma$ & $5.61(21)$ & $0.1355(26)$ \\
		\midrule
		$\rho \to \eta \gamma$ & $44.2(3.1)$ & $1.58(6)$ \\
		$\omega \to \eta \gamma$ & $3.91(35)$ & $0.449(20)$ \\
		$\phi \to \eta \gamma$ & $55.3(1.1)$ & $0.691(7)$ \\
		\midrule
		$\eta' \to \rho \gamma$ & $55.5(1.9)$ & $1.299(23)$ \\
		$\eta' \to \omega \gamma$ & $4.74(20)$ & $0.401(9)$ \\
		$\phi \to \eta' \gamma$ & $0.264(9)$ & $0.712(12)$ \\
		\bottomrule
	\end{tabular}
	\caption{The normalizations $\lvert \F_{V P}(0) \rvert$ at the real-photon point obtained from \autoref{eq:width_VPgamma} and phenomenological input determined from Ref.~\cite{ParticleDataGroup:2022pth}; see also \autoref{tab:constants}.}
	\label{tab:couplings_pheno}
\end{table}
The construction of the $\CC$-even decay amplitudes for
\begin{align}
	\eta^{(\prime)}(P) 
    &\to \pi^0(p_0) \ell^+(p_+) \ell^-(p_-), \notag \\
	\eta'(P) 
    &\to \eta(p_0) \ell^+(p_+) \ell^-(p_-),
\end{align}
where $\ell = e, \mu$, is based on the assumption that the underlying $\eta^{(\prime)} \to \pi^0 \gamma^* \gamma^*$ and $\eta' \to \eta \gamma^* \gamma^*$ amplitudes are dominated by the exchange of the vector mesons $V = \rho, \omega, \phi$, $\phi \equiv \phi(1020)$; see \autoref{fig:fd_process}.
For our analysis, we define the \LN{Mandelstam} variables $s = (p_+ + p_-)^2$, $t = (p_- + p_0)^2$, and $u = (p_+ + p_0)^2$, which describe the invariant mass squares of the lepton pair and the lepton--pseudoscalar subsystems, respectively; they fulfill the relation $\Sigma = s + t + u = \massIn^2 + \massOut^2 + 2m_\ell^2$.
The relevant vector-to-pseudoscalar transition form factors $\F_{V P}(q^2)$ are defined according to
\begin{align}\label{eq:matrix_element_VPgamma}
	\langle P(p) | j_\mu(0) | V(p_V) \rangle 
    = e \, \eps_{\mu \nu \alpha \beta} \eps^\nu(p_V) p^\alpha q^\beta \F_{V P}(q^2),
\end{align}
where $j_\mu = e(2 \bar u\gamma_\mu u - \bar d \gamma_\mu d -\bar s \gamma_\mu s)/3$ denotes the electromagnetic current, and $q = p_V - p$.
The normalizations $\lvert \F_{V P}(0) \rvert$ at the real-photon point can be derived from phenomenological input in a straightforward manner,
\begin{align}\label{eq:width_VPgamma}
	\Gamma(V \to P \gamma) 
    &= \frac{\alpha (M_V^2 - M_P^2)^3}{24 M_V^3} \lvert \F_{V P}(0) \rvert^2, \notag \\
	\Gamma(P \to V \gamma) 
    &= \frac{\alpha (M_P^2 - M_V^2)^3}{8 M_P^3} \lvert \F_{V P}(0) \rvert^2,
\end{align}
where $\alpha = e^2/(4\pi)$ is the fine-structure constant, leading to \autoref{tab:couplings_pheno} with input from Ref.~\cite{ParticleDataGroup:2022pth}.

Using \autoref{eq:matrix_element_VPgamma} and summing over the $t$- and $u$-channel diagrams shown in \autoref{fig:fd_process} as well as $V = \rho, \omega, \phi$, we find the amplitude $\M \equiv \M(\eta^{(\prime)} \to [\pi^0/\eta] \ell^+ \ell^-)$ to be
\begin{widetext}
\begin{align}\label{eq:amplitude}
	\M 
    &= \iu \frac{\alpha^2}{\pi^2} \sum_V \int \dfk \, g^{\beta \tilde{\beta}} \eps_{\mu \nu \alpha \beta} \eps_{\tilde{\mu} \tilde{\nu} \tilde{\alpha} \tilde{\beta}} P^\alpha k^\mu \big(P^{\tilde{\alpha}} k^{\tilde{\mu}} - P^{\tilde{\alpha}} l^{\tilde{\mu}} + k^{\tilde{\alpha}} l^{\tilde{\mu}}\big) P_V^\BW\big((P - k)^2\big) P_\gamma(k^2) P_\gamma\big((l - k)^2\big) \notag \\
	&\qquad \qquad \times \F_{V \eta^{(\prime)}}(k^2) \F_{V [\pi^0/\eta]}\big((l-k)^2\big) \bar{u}_s \bigg[ \gamma^{\tilde{\nu}} \frac{\slashed{k} - \slashed{p}_+ + m_\ell}{(k - p_+)^2 - m_\ell^2} \gamma^\nu + \gamma^\nu \frac{\slashed{p}_- - \slashed{k} + m_\ell}{(p_- - k)^2 - m_\ell^2} \gamma^{\tilde{\nu}} \bigg] v_r,
\end{align}
\end{widetext}
with $\bar{u}_s \equiv \bar{u}_s(p_-)$ and $v_r \equiv v_r(p_+)$.
Here, we defined $l = p_+ + p_-$ and the [\LN{Breit}--\LN{Wigner} (BW)] propagators 
\beq\label{eq:propagators}
	P_V^\BW(q^2) 
    = \frac{1}{q^2 - M_V^2 + \iu M_V \Gamma_V},
	\,\,\,
	P_\gamma(q^2) 
    = \frac{1}{q^2 + \iu \eps},
\eeq
where $M_V$ is the mass of the respective vector meson and $\Gamma_V$ its width.
Due to their narrowness, a constant-width approximation is well justified for the $\omega$ and $\phi$, whereas the broad $\rho$ meson necessitates an energy-dependent width to avoid sizable unphysical imaginary parts below threshold. 
We will implement such a parameterization for the $\rho$ in \autoref{sec:spectral_rep}, where we will use a dispersively improved BW propagator.
Our final results will be quoted for both a variant with constant widths for all vector mesons (CW) and a variant that instead employs an energy-dependent width for the $\rho$ (VW).

For the eventual computations, it will turn out useful to apply the \LN{Dirac} equation and make the replacements
\begin{align}\label{eq:Dirac_eq}
	\bar{u}_s \gamma^{\tilde{\nu}} (\slashed{k} - \slashed{p}_+ + m_\ell) \gamma^\nu v_r 
    &= \bar{u}_s (\gamma^{\tilde{\nu}} \slashed{k} \gamma^\nu - 2p_+^\nu \gamma^{\tilde{\nu}}) v_r, \notag \\
	\bar{u}_s \gamma^\nu (\slashed{p}_- - \slashed{k} + m_\ell) \gamma^{\tilde{\nu}} v_r 
    &= \bar{u}_s (2p_-^\nu \gamma^{\tilde{\nu}} - \gamma^\nu \slashed{k} \gamma^{\tilde{\nu}}) v_r
\end{align}
in \autoref{eq:amplitude}.

\begin{figure}[t]
	\centering
	\includegraphics[scale=0.9]{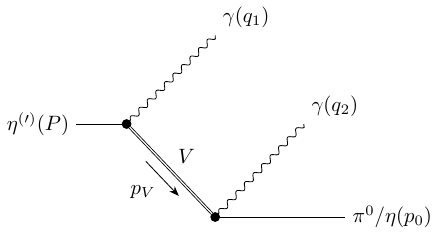}
	\\[1cm]
	\includegraphics[scale=0.9]{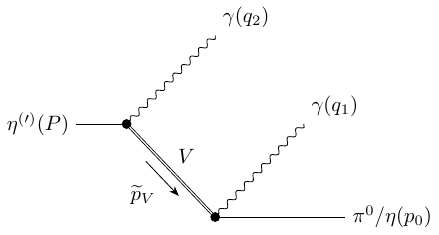}
	\caption{The two diagrams contributing to the two-photon decay $\eta^{(\prime)} \to [\pi^0/\eta] \gamma \gamma$, which are related via $q_1 \leftrightarrow q_2$.}
	\label{fig:fd_two_photon_decay}
\end{figure}
The branching ratios of the semileptonic decays are commonly normalized to the two-photon analogs
\begin{align}
    \eta^{(\prime)}(P) 
    &\to \pi^0(p_0) \gamma(q_1) \gamma(q_2), \notag \\
    \eta'(P)
    &\to \eta(p_0) \gamma(q_1) \gamma(q_2);
\end{align}
see also \autoref{fig:fd_two_photon_decay}.
For these decays, we define the \LN{Mandelstam} variables\footnote{%
    Note that the \LN{Mandelstam} variable $s = (P - p_0)^2$ is identical in the semileptonic and the diphoton case.}
\beq\label{eq:Mandelstam_gamma}
    s 
    = (q_1 + q_2)^2, 
    \quad 
    t_\gamma 
    = (q_2 + p_0)^2, 
    \quad 
    u_\gamma 
    = (q_1 + p_0)^2,
\eeq 
which fulfill $\Sigma_\gamma = s + t_\gamma + u_\gamma = \massIn^2 + \massOut^2$, and denote the corresponding helicity amplitudes by $\Hel_{\lambda\lambda'}$,
\begin{align}\label{eq:def_hel_amp}
    &\langle \gamma(q_1,\lambda) \gamma(q_2,\lambda') | \,S\, | \eta^{(\prime)}(P) [\pi^0/\eta](p_0) \rangle \notag \\
    &= \iu (4\pi\alpha) (2\pi)^4 \delta^{(4)}(P + p_0 - q_1 - q_2)\, e^{\iu (\lambda - \lambda') \varphi} \Hel_{\lambda\lambda'}. 
\end{align}
Here, $\lambda^{(\prime)}$ are the helicities of the photons and we factored out the dependence on the electric charge $(4\pi\alpha)$ and the azimuthal angle $\varphi$ for convenience.
Using \autoref{eq:matrix_element_VPgamma} and the normalization of the form factors, $\lvert C_{VP\gamma} \rvert = \lvert \F_{VP}(0) \rvert$, as will be introduced in \autoref{sec:form_factors}, we express the VMD helicity amplitudes as
\begin{align}\label{eq:amplitude_gamma_gamma}
    \Hel_{\lambda\lambda'}
    &= {\eps_{\lambda}^{\alpha_1}}^*(q_1) {\eps_{\lambda'}^{\alpha_2}}^*(q_2) \sum_V C_{V \eta^{(\prime)} \gamma}C_{V[\pi^0/\eta]\gamma} \\
    &\qquad\qquad \times \big[ P_V^\BW(t_\gamma) \Hel_{\alpha_1 \alpha_2}^t + P_V^\BW(u_\gamma) \Hel_{\alpha_1 \alpha_2}^u \big], \notag
\end{align}
where $\eps_{\lambda}^*(q_i)$ denote the polarization vectors of the outgoing photons and 
\begin{align}
    \Hel_{\alpha_1 \alpha_2}^t 
    &= g^{\mu_1 \mu_2} \eps_{\mu_1 \nu_1 \alpha_1 \beta_1} \eps_{\mu_2 \nu_2 \alpha_2 \beta_2} p_V^{\nu_1} q_1^{\beta_1} p_0^{\nu_2} q_2^{\beta_2}, \notag \\
    \Hel_{\alpha_1 \alpha_2}^u 
    &= g^{\mu_1 \mu_2} \eps_{\mu_1 \nu_1 \alpha_1 \beta_1} \eps_{\mu_2 \nu_2 \alpha_2 \beta_2} p_0^{\nu_1} q_1^{\beta_1} \widetilde{p}_V^{\nu_2} q_2^{\beta_2},
\end{align}
with $p_V = q_2 + p_0$ and $\widetilde{p}_V = q_1 + p_0$ the momenta of the intermediate vector mesons.

%---------------------------------------------------------------------------------------------------
\section{Form factors}
\label{sec:form_factors}
%---------------------------------------------------------------------------------------------------
\begin{figure}[t]
	\centering
	\includegraphics{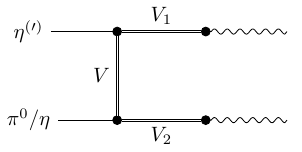}
	\caption{The modeling of the two-photon decay mechanism in the VMD framework via two vector mesons $V_1$ and $V_2$.
    Constraints on $(V_1,V_2)$ in dependence of the initial and final state as well as $V$ are given in \autoref{tab:VMD_allowed}.}
	\label{fig:fd_vmd}
\end{figure}
In order to parameterize the form factors $\F_{V P}(q^2)$, we use the VMD framework.
As a consequence, the photon couplings at the $V P \gamma^*$ vertices of the diagrams in \autoref{fig:fd_process} are mediated via two intermediate vector mesons $V_1$ and $V_2$; see \autoref{fig:fd_vmd}. 
We will construct two distinct such models: a monopole (MP) parameterization with $V_i = \rho, \omega, \phi$ and a dipole (DP) ansatz with $V_i = \rho^{(\prime)}, \omega^{(\prime)}, \phi^{(\prime)}$, $\rho' \equiv \rho^0(1450)$, $\omega' \equiv \omega(1420)$, and $\phi' \equiv \phi(1680)$, that ensures the expected high-energy behavior of the form factors~\cite{Farrar:1975yb,Vainshtein:1977db,Lepage:1979zb,Lepage:1980fj,Chernyak:1983ej}.
For reference, we also include a model calculation with constant form factors, \Lat{i.e.}, a point-like (PL) interaction, which closely resembles the parameterization of Ref.~\cite{Escribano:2020rfs}.

The conservation of isospin---and thus $\GG$ parity combined with $\CC$---imposes constraints on $V_1$ and $V_2$ in dependence on the initial and final states as well as the $t$- or $u$-channel vector meson $V$.
However, some of the couplings, namely $\eta^{(\prime)} \omega \phi^{(\prime)}$, $\eta^{(\prime)} \phi \omega^{(\prime)}$, $\pi^0 \rho \phi^{(\prime)}$, and $\pi^0 \phi \rho^{(\prime)}$, are, although isospin-allowed, vanishing under the assumption of $\Uthree$ flavor symmetry and ideally mixed vector-meson multiplets; see Appendix~\ref{appx:u3}.
Since the contribution of $V = \phi$ would otherwise vanish entirely for $\eta^{(\prime)} \to \pi^0 \ell^+ \ell^-$, we nonetheless include the \LN{Okubo}--\LN{Zweig}--\LN{Iizuka}-suppressed (OZI-suppressed)~\cite{Okubo:1963fa,Zweig:1964jf,Iizuka:1966fk} couplings $\pi^0 \phi \rho^{(\prime)}$ in our calculations; the remaining vector mesons $V_i$ are collected in \autoref{tab:VMD_allowed}.

%---------------------------------------------------------------------------------------------------
\subsection{Monopole model}
\label{sec:monopole}
%---------------------------------------------------------------------------------------------------
The MP model only takes the lowest-lying vector mesons $\rho$, $\omega$, and $\phi$ into account, so that the form factors are parameterized according to
\beq\label{eq:form_factors_MP_CW}
	\F_{V P}(q^2) 
    = C_{V P \gamma} M_{V_i}^2 P_{V_i}^\BW(q^2),
\eeq
with the assignments of $V_i \in \{\rho, \omega, \phi\}$ according to \autoref{tab:VMD_allowed}.
Here, we assume $\lvert C_{V P \gamma} \rvert = \lvert \F_{V P}(0) \rvert$ at the real-photon point, see \autoref{tab:couplings_pheno}, which determines the coupling constants $C_{V P \gamma}$ up to an overall phase.
This assumption omits corrections due to the constant, non-zero widths in the BW propagators, which are negligible for $V = \omega, \phi$ but potentially significant for $V = \rho$.\footnote{%
    Note that $P_V^\BW(0) \simeq -1/M_V^2$, so that $\F_{V P}(0) = -C_{V P \gamma}$, which, however, corresponds to an unobservable overall phase.}
Since the energy-dependent width of the $\rho$ meson will be chosen to have the proper threshold behavior, these complications only exist for the variant CW but not for VW.
All coupling constants are assumed to be real in the following.
In order to fix the relative signs between them, we resort to $\Uthree$ flavor symmetry and analyses of $e^+ e^- \to 3 \pi$ and $e^+ e^- \to \pi \gamma$~\cite{Hoferichter:2014vra,Hoferichter:2019mqg,Hoid:2020xjs}; see Appendix~\ref{appx:u3}.
Without loss of generality, we adopt a positive sign for the coupling $C_{\rho \eta \gamma}$ and establish the consistent sign convention collected in \autoref{tab:couplings_signs}.

%---------------------------------------------------------------------------------------------------
\subsection{Dipole model}
\label{sec:dipole}
%---------------------------------------------------------------------------------------------------
\begin{table}[t]
	\centering
	\begin{tabular}{l  c  l  l  l  c  l  l  l}
	\toprule
		& & \multicolumn{3}{c}{$V \pi^0 \gamma$} & & \multicolumn{3}{c}{$V \eta^{(\prime)} \gamma$} \\
		\midrule
    	$V$ & & $\rho$ & $\omega$ & $\phi$ & & $\rho$ & $\omega$ & $\phi$ \\
        \\[-0.35cm]
        $V_i$ & & $\omega^{(\prime)}$ & $\rho^{(\prime)}$ & $\rho^{(\prime)}$ & & $\rho^{(\prime)}$ & $\omega^{(\prime)}$ & $\phi^{(\prime)}$ \\
		\bottomrule
	\end{tabular}
    \caption{The constraints on the vector mesons $V_i$ of \autoref{fig:fd_vmd} in dependence of $V$ derived from isospin conservation and $\Uthree$ flavor symmetry with ideally mixed vector-meson multiplets. 
    We include the OZI-suppressed couplings $\phi \pi^0 \rho^{(\prime)}$; see text and Appendix~\ref{appx:u3} for more information.}
	\label{tab:VMD_allowed}
\end{table}
Given that the asymptotic behavior of the vector-to-pseudoscalar transition form factors is expected to be $\F_{V P}(q^2) \propto q^{-4}$~\cite{Farrar:1975yb,Vainshtein:1977db,Lepage:1979zb,Lepage:1980fj,Chernyak:1983ej}, we can additionally include the next-higher multiplet of vector mesons, $\rho'$, $\omega'$, and $\phi'$, to achieve this property by tuning a free parameter $\eps_V$.\footnote{%
    Data both on $e^+ e^- \to \omega \pi^0$~\cite{Achasov:2016zvn} and $e^+ e^- \to \rho^0 \eta$~\cite{BaBar:2007qju,BaBar:2018erh,Holz:2015tcg} suggest that the required cancellation indeed largely occurs between the contributions of the two lowest vector states, $\rho$ and $\rho'$ in those cases.}
For the DP model, we thus make the ansatz
\begin{align}\label{eq:form_factors_DP_CW}
	\widetilde{\F}_{V P}(q^2) 
    &= C_{V P \gamma} \big[ (1 - \eps_{V_i}) M_{V_i}^2 P_{V_i}^\BW(q^2) \notag \\
    &\qquad \qquad + \eps_{V_i} M_{V_i'}^2 P_{V_i'}^\BW(q^2) \big],
\end{align}
where we assume the excited vector states to couple according to the exact same symmetry restrictions as the ground-state multiplet; \Lat{cf.}\ \autoref{tab:VMD_allowed}.
Here, $P_{V'}^\BW(q^2)$ is defined as in \autoref{eq:propagators}, with $M_{V'}$ and $\Gamma_{V'}$ the mass and width of the respective excited vector meson.
Due to the large widths of the excited vector mesons, a constant-width approximation leads to a rather poor description of these mesons, however.
We will therefore, analogously to the $\rho$, construct dispersively improved BW propagators for $\rho'$, $\omega'$, and $\phi'$ based on energy-dependent widths in \autoref{sec:spectral_rep}, leading to replacements of the kind $P_{V'}^\BW(q^2) \to P_{V'}^\disp(q^2)$.
Similarly to the MP, our final results for the DP will be quoted for both the variant CW with constant widths for all vector mesons and the variant VW, \Lat{i.e.}, using constant widths for the $\omega$ and $\phi$ but energy-dependent ones for $\rho^{(\prime)}$, $\omega'$, and $\phi'$.
The form factors in \autoref{eq:form_factors_DP_CW} are assumed to be normalized such that $\widetilde{F}_{V P}(0) = -C_{V P \gamma}$, which, as for the MP, holds up to potential corrections due to the constant widths in the propagators. 
In order to achieve the desired high-energy behavior, the free parameter needs to be chosen as $\eps_V = M_V^2/(M_V^2 - M_{V'}^2)$.

%---------------------------------------------------------------------------------------------------
\subsection{Spectral representation}
\label{sec:spectral_rep}
%---------------------------------------------------------------------------------------------------
While the variant CW has its own \Lat{raison d'être} as a simple approximate description, the large widths of the mesons $\rho^{(\prime)}$, $\omega'$, and $\phi'$ actually require an energy-dependent parameterization to avoid significant unphysical imaginary parts below threshold.\footnote{%
    In principle, such unphysical imaginary parts could be avoided for the $\rho$ exchange by reconstructing the latter in terms of dispersion relations for $\gamma^{(*)} \pi \to \pi \pi$~\cite{Hoferichter:2012pm,Hoferichter:2017ftn,Niehus:2021iin} and $\eta^{(\prime)} \to \pi \pi \gamma^{(*)}$~\cite{Stollenwerk:2011zz,Kubis:2015sga,Holz:2015tcg,Holz:2022hwz}; \Lat{cf.}\ also Refs.~\cite{Colangelo:2014pva,Hoferichter:2019nlq}.
    We here refrain from further refining the amplitude in such a way.}
In this section, we construct these energy-dependent widths; to ensure the correct analytic properties when inserting the form factors into the amplitude, \autoref{eq:amplitude}, we will furthermore introduce dispersively improved variants~\cite{Crivellin:2022gfu} of the form factors that contain a $\rho^{(\prime)}$-, $\omega'$-, or $\phi'$-meson propagator, which lay the foundation for the variant VW in both the MP and DP model.
\begin{table}[t]
	\centering
    {
    \setlength{\tabcolsep}{10.0pt}
	\begin{tabular}{c  c  c}
	\toprule
        $C_{\rho \pi^0 \gamma}$ & $C_{\omega \pi^0 \gamma}$ & $C_{\phi \pi^0 \gamma}$ \\
        \\[-0.35cm]
        $+$ & $+$ & $-$ \\
		\midrule
		$C_{\rho \eta \gamma}$ & $C_{\omega \eta \gamma}$ & $C_{\phi \eta \gamma}$ \\
        \\[-0.35cm]
        $+$ & $+$ & $-$ \\
		\midrule
        $C_{\rho \eta' \gamma}$ & $C_{\omega \eta' \gamma}$ & $C_{\phi \eta' \gamma}$ \\
        \\[-0.35cm]
        $+$ & $+$ & $+$ \\
		\bottomrule
	\end{tabular}
    }
	\caption{The signs $\sgn[C_{V P \gamma}]$ of the couplings constants defined in \autoref{eq:form_factors_MP_CW}.
    Here, we fixed the global sign of $C_{\rho \eta \gamma}$ to be positive; see Appendix~\ref{appx:u3} for details.}
	\label{tab:couplings_signs}
\end{table}

For the $\rho$ meson, we will use the energy-dependent width~\cite{Zanke:2021wiq}
\begin{align}
	&\Grho(q^2) 
    = \theta(q^2 - 4\MpiC^2) \frac{\gamma_{\rho \to \pi^+ \pi^-}(q^2)}{\gamma_{\rho \to \pi^+ \pi^-}(\Mrho^2)} f(q^2) \Grho, \notag \\
	&\gamma_{\rho \to \pi^+ \pi^-}(q^2) 
    = \frac{(q^2 - 4\MpiC^2)^{3/2}}{q^2},
\end{align}
where the so-called barrier factor~\cite{COMPASS:2015gxz,VonHippel:1972fg}
\beq
	f(q^2) 
    = \frac{\sqrt{q^2}}{\Mrho} \frac{\Mrho^2 - 4\MpiC^2 + 4p_R^2}{q^2 - 4\MpiC^2 + 4p_R^2}, 
\eeq
$p_R = 202.4 \MeV$, has been introduced to ensure convergence of the superconvergence relations evaluated in \autoref{eq:sc_relations} below. 
We calculate the dispersive $\rho$ propagator via
\begin{align}\label{eq:propagator_disp}
	P_V^\disp(q^2) 
    &= -\frac{1}{\pi} \int_{s_\thr}^\infty{\dff x \, \frac{\Im[P_V^\BW(x)]}{q^2 - x + \iu \eps}}, \notag \\
	\Im[P_V^\BW(x)] 
    &= \frac{-\sqrt{x} \, \Gamma_V(x)}{(x - M_V^2)^2 + x \Gamma_V(x)^2},
\end{align}
where $s_\thr = 4\MpiC^2$ is the threshold for $\rho \to \pi^+ \pi^-$.
The spectral representations of the form factors $\F_{V P}(q^2)$ for $V P \in \{\rho \eta^{(\prime)}, \omega \pi^0, \phi \pi^0\}$ are thus given by
\beq
	\widehat{\F}_{V P}(q^2) 
    = \frac{C_{V P \gamma}}{N_\rho} \Mrho^2 P_\rho^\disp(q^2),
\eeq
where the normalization constant
\beq
	N_\rho 
    = -\Mrho^2 P_\rho^\disp(0) 
    \approx 0.898 % 0.898308 % KEEP!
\eeq
is introduced in order to retain $\widehat{\F}_{V P}(0) = -C_{V P \gamma}$, \Lat{i.e.}, to ensure that the coupling constants have the same meaning in the original and the dispersively improved VMD parameterization.
For reasons of consistency, we also replace the $\rho$ propagator in the left-hand cuts, $P_\rho^\BW(q^2)$ in \autoref{eq:amplitude}, by a dispersively improved variant, \Lat{i.e.},
\beq
	P_\rho^\BW(q^2) 
    \to \frac{1}{N_\rho^\text{LHC}} P_\rho^\disp(q^2),
\eeq
where the normalization constant
\beq
	N_\rho^\text{LHC} 
    = \iu \Mrho \Grho P_\rho^\disp(\Mrho^2) 
    \approx 1 
\eeq
is introduced in order to retain $P_\rho^\BW(\Mrho^2) = 1/(\iu \Mrho \Grho)$, in line with the VMD assumption.\footnote{%
    We ignore the fact that the $\rho$ pole in the complex plane does not exactly agree with the \LN{Breit}--\LN{Wigner} parameters.}
With these conventions, we will drop the distinction between $\F_{V P}(q^2)$ and $\widehat{\F}_{V P}(q^2)$ in the following, and it will always be clear from context which representation is used.

\begin{table}[t]
	\centering
	\begin{tabular}{c  c  c  c  c}
	\toprule
		$\eps_\rho$ & $(-0.47)\substack{-0.07\\+0.06}$ & \phantom{...} & $\widetilde{N}_\rho$ & $0.99\substack{+0.04\\-0.03}$ \\
        \\[-0.35cm]
		$\eps_\omega$ & $(-0.43)\substack{-0.25\\+0.16}$ & & $\widetilde{N}_\omega$ & $1.10\substack{+0.17\\-0.10}$ \\
        \\[-0.35cm]
		$\eps_\phi$ & $(-0.42)\substack{-0.08\\+0.06}$ & & $\widetilde{N}_\phi$ & $1.03\substack{+0.05\\-0.04}$ \\
		\bottomrule
	\end{tabular}
	\caption{The values of the parameter $\eps_V$ derived from the superconvergence relations, \autoref{eq:eps_scr}, and the normalization constants of \autoref{eq:normalization_dipole}.
    Here, tiny imaginary parts in the normalization constants have been neglected.
	The uncertainties refer to the variations of $\GrhoPrime$, $\GomegaPrime$, and $\GphiPrime$, see \autoref{tab:constants}, and are omitted in the subsequent analysis.}
	\label{tab:eps_and_N}
\end{table}
For the dipole variant, the widths of the excited vector mesons $\rho', \omega', \phi'$ are modeled using the dominant quasi-two-particle thresholds.
We condense the decays $\rho' \to \omega \pi$, $\omega' \to \rho \pi$, and $\phi' \to K^* \bar K$, $K^* \equiv {K^*}(892)$, in the notation $V' \to V P$, such that
\begin{align}
	\Gamma_{V'}(q^2) 
    &= \theta\big(q^2 - (M_V + M_P)^2\big) \frac{\gamma_{V' \to V P}(q^2)}{\gamma_{V' \to V P}(M_{V'}^2)} \Gamma_{V'}, \notag \\
	\gamma_{V' \to V P}(q^2) 
    &= \frac{\lambda(q^2,M_V^2,M_P^2)^{3/2}}{(q^2)^{3/2}},
\end{align}
where $\lambda(x,y,z) = x^2 + y^2 + z^2 - 2xy - 2xz - 2yz$ is the \LN{K\"all\'en} function.
Here, we disregard any distinction between the various charge channels and use the neutral masses for numerical evaluation.
The dispersive $\rho'$, $\omega'$, and $\phi'$ propagators and spectral functions are defined similarly to \autoref{eq:propagator_disp}, with $s_\thr = (M_V + M_P)^2$ for the thresholds.
In analogy to \autoref{eq:form_factors_DP_CW}, the dipole form factors read
\begin{align}\label{eq:form_factor_DP_VW}
	\widetilde{\F}_{V P}(q^2) 
    &= \frac{C_{V P \gamma}}{\widetilde{N}_{V_i}} \big[ (1 - \eps_{V_i}) M_{V_i}^2 P_{V_i}(q^2) \notag \\
    &\qquad \qquad \,\, + \eps_{V_i} M_{V_i'}^2 P_{V_i'}(q^2) \big],
\end{align}
where the simplifying assumption of constant widths for $\omega$ and $\phi$ propagators is always implicitly understood, with $P_{V_i'}(q^2) \in \{P_{V_i'}^\BW(q^2),P_{V_i'}^\disp(q^2)\}$.
Here, we introduced the normalization constants
\beq\label{eq:normalization_dipole}
	\widetilde{N}_V
    = -\big[ (1 - \eps_V) M_V^2 P_V(0) + \eps_V M_{V'}^2 P_{V'}(0) \big],
\eeq
which, once more, ensure $\widetilde{\F}_{V P}(0) = -C_{V P \gamma}$.
The parameters $\eps_V$ have to be tuned differently in the dispersively improved variant, namely via the superconvergence relations
\begin{align}\label{eq:sc_relations}
    0 
    &= (1 - \eps_V) M_V^2 P_V^0 + \eps_V M_{V'}^2 P_{V'}^0, \\
	P_V^0 
    &=
    \begin{cases}
        \phantom{-} 1 \,, 
        & V = \omega, \phi, 
        \\[2mm]
        - \displaystyle{\frac{1}{\pi} \int_{s_\thr}^\infty} \dff x \, \Im[P_V^\BW(x)] \,, 
        & V=\rho^{(\prime)}, \omega', \phi', \notag
    \end{cases}
\end{align}
such that the terms of $\Order(1/q^2)$ in the form factors cancel.
We collect the numerical results for
\beq\label{eq:eps_scr}
	\eps_V 
    = \frac{M_V^2 P_V^0}{M_V^2 P_V^0 - M_{V'}^2 P_{V'}^0} 
\eeq
and $\widetilde{N}_V$ in \autoref{tab:eps_and_N}, where we include the uncertainties due to the large errors on $\Gamma_{V'}$; in the following, their effect is, however, assumed to be insignificant and thus discarded.

%---------------------------------------------------------------------------------------------------
\section{Observables}
\label{sec:observables}
%---------------------------------------------------------------------------------------------------
The phenomenological analysis in this article will be performed in terms of doubly- and singly-differential decay widths as well as integrated branching ratios.
We define $\nu = t - u$ for the \LN{Mandelstam} variables $t$ and $u$, in terms of which the twofold differential decay width $\dff\Gamma \equiv \dff\Gamma(\eta^{(\prime)} \to [\pi^0/\eta] \ell^+ \ell^-)$ is given by~\cite{ParticleDataGroup:2022pth}
\beq\label{eq:DW}
	\dff\Gamma 
    = \frac{1}{(2\pi)^3} \frac{1}{64 \massIn^3} \lvert \overline{\M} \rvert^2 \dff s \, \dff\nu.
\eeq
Here, $\lvert \overline{\M} \rvert^2$ is the spin-summed square of the amplitude, \autoref{eq:amplitude}, and the integration region is bounded by the available phase space
\begin{align}\label{eq:Mandelstam_limits}
    s &\in [4m_\ell^2, (\massIn - \massOut)^2], \notag \\
    \nu &\in [-\nu_\max, \nu_\max], 
    \quad
    \nu_\max
    = \sigma(s) \sqrt{\lambda(s)},
\end{align}
with
\beq
    \sigma(s) 
    = \sqrt{1 - \frac{4m_\ell^2}{s}}, \quad \lambda(s) \equiv \lambda(s,\massIn^2,\massOut^2).
\eeq
The singly-differential decay width $\dff\Gamma/\dff s$ follows from an integration of \autoref{eq:DW} over $\nu$ and the branching ratio
\beq\label{eq:BR}
    \BR(\eta^{(\prime)} \to [\pi^0/\eta] \ell^+ \ell^-) 
    = \frac{\Gamma}{\GammaIn}
\eeq
is obtained after performing the full three-body phase-space integration, \Lat{i.e.}, by also integrating over $s$.

\begin{table}[t]
	\centering
	\begin{tabular}{l  r  r  r}
	\toprule
		& $C_\rho/\GeV^{-2}$ & $C_\omega/\GeV^{-2}$ & $C_\phi/\GeV^{-2}$ \\
		\midrule
        $\eta \to \pi^0 \ell^+ \ell^-$ & $1.16(11)$ & $1.05(5)$ & $0.0936(20)$ \\
		$\eta' \to \pi^0 \ell^+ \ell^-$ & $0.95(8)$ & $0.937(26)$ & $-0.0965(25)$ \\ 
		$\eta' \to \eta \ell^+ \ell^-$ & $2.05(8)$ & $0.180(9)$ & $-0.492(10)$ \\ 
		\bottomrule
	\end{tabular}
	\caption{Numerical values of the coupling constants defined in \autoref{eq:PV_result_amplitude} for the different processes.}
	\label{tab:couplings_CV}
\end{table}
In order to calculate $\lvert \overline{\M} \rvert^2$, we perform a PV decomposition of \autoref{eq:amplitude} with \Lat{FeynCalc}~\cite{Shtabovenko:2016sxi,Shtabovenko:2020gxv,Mertig:1990an} after inserting explicit expressions for the form factors.
For both the MP and DP model and in both variants CW and VW, this results in an expression of the generic form
\begin{align}\label{eq:PV_result_amplitude}
 	\M 
    &= 16\pi^2 \alpha^2 \big[ \M_\QED^{uv} \M_\had^{uv} + \M_\QED^{u0v} \M_\had^{u0v} \big], \notag \\
    \M_\QED^{uv} 
    &= m_\ell \, \bar{u}_s v_r,
    \quad
 	\M_\QED^{u0v}
    = \bar{u}_s \slashed{p}_0 v_r, \\
	\M_\had^{u(0)v} 
    &= \sum_V C_V \M_V^{u(0)v},
    \quad
    C_V
    = C_{V \eta^{(\prime)} \gamma} C_{V [\pi^0/\eta] \gamma}, \notag
\end{align}
where the quantities $\M_V^{u(0)v}$ account for the different vector-meson contributions in the result of the PV decomposition, \Lat{cf.}\ the sum in \autoref{eq:amplitude}; they amount to cumbersome expressions containing PV functions.\footnote{%
    These expressions are attached as a text file to this article~\cite{SupplementalMaterialBoth}.}
The numerical values of the process-specific coupling constants $C_V$ are provided in \autoref{tab:couplings_CV}.
Upon squaring and spin-summing, the above amplitude leads to
\begin{align}\label{eq:PV_result_amplitude_squared}
    \lvert \overline{\M} \vert^2 
    &= 256 \pi^4 \alpha^4 \big[ C_\rho^2 \, \lvert \overline{\M}_{\rho,\rho} \rvert^2 + C_\omega^2 \, \lvert \overline{\M}_{\omega,\omega} \rvert^2 + C_\phi^2 \, \lvert \overline{\M}_{\phi,\phi} \rvert^2 \notag \\
    &\qquad\qquad\quad + C_\rho C_\omega \, \lvert \overline{\M}_{\rho,\omega} \rvert^2 + C_\rho C_\phi \, \lvert \overline{\M}_{\rho,\phi} \rvert^2 \notag \\
    &\qquad\qquad\quad + C_\omega C_\phi \, \lvert \overline{\M}_{\omega,\phi} \rvert^2 \big],
\end{align}
where we defined
\begin{align}
    \lvert \overline{\M}_{V,V} \rvert^2 
    &= \lvert \overline{\M}_\QED^{uv} \rvert^2 \lvert \M_V^{uv} \rvert^2 + \lvert \overline{\M}_\QED^{u0v} \rvert^2 \lvert \M_V^{u0v} \rvert^2 \notag \\
    &\quad + 2 \, \overline{\M}_\QED^{uv,u0v} \Re\big[ \M_V^{uv} {\M_V^{u0v}}^* \big], \notag \\
    \lvert \overline{\M}_{V_1,V_2} \rvert^2
    &= 2 \, \lvert \overline{\M}_\QED^{uv} \rvert^2 \Re\big[ \M_{V_1}^{uv} {\M_{V_2}^{uv}}^* \big] \notag \\
    &\quad + 2 \, \lvert \overline{\M}_\QED^{u0v} \rvert^2 \Re\big[ \M_{V_1}^{u0v} {\M_{V_2}^{u0v}}^* \big] \\
    &\quad + 2 \, \overline{\M}_\QED^{uv,u0v} \Re\big[ \M_{V_1}^{uv} {\M_{V_2}^{u0v}}^* + \M_{V_1}^{u0v} {\M_{V_2}^{uv}}^* \big] \notag
\end{align}
for $V_1 \neq V_2$, with
\begin{align}
    \lvert \overline{\M}_\QED^{uv} \rvert^2
    &= 2 m_\ell^2 (s - 4m_\ell^2), 
    &
    \lvert \overline{\M}_\QED^{u0v} \rvert^2
    &= \frac{1}{2} \big[ \lambda(s) - \nu^2 \big], \notag \\
    \overline{\M}_\QED^{uv,u0v} 
    &= - 2m_\ell^2 \nu.
\end{align}

Similarly to the semileptonic decays, the branching ratio of the two-photon analogs is defined by
\beq\label{eq:BR_gamma_gamma}
    \BR(\eta^{(\prime)} \to [\pi^0/\eta] \gamma \gamma) 
    = \frac{\Gamma_\gamma}{\GammaIn},
\eeq
where $\Gamma_\gamma \equiv \Gamma(\eta^{(\prime)} \to [\pi^0/\eta] \gamma \gamma)$ and
\beq\label{eq:DW_gamma_gamma}
    \dff\Gamma_\gamma
    = \frac{1}{(2\pi)^3} \frac{(4\pi\alpha)^2}{64 \massIn^3} \lvert \overline{\Hel} \rvert^2 \dff s\, \dff \nu_\gamma,
\eeq
with the phase space bounded by
\begin{align}
    s &\in [0,(\massIn - \massOut)^2], \notag \\
    \nu_\gamma &\in [-\nu_\gamma^\max, \nu_\gamma^\max], 
    \quad
    \nu_\gamma^\max
    = \sqrt{\lambda(s)}.
\end{align}
Due to the indistinguishability of the two photons in the final state, an additional factor of $1/2$ has to be taken into account upon integration.
From \autoref{eq:amplitude_gamma_gamma}, one finds the polarization-summed amplitude squared
\begin{align}\label{eq:amplitude_squared_gamma_gamma}
    \lvert \overline{\Hel} \rvert^2 
    &= \frac{1}{8} \bigg[ \sum_V C_V^2 \Big( \big\lvert P_V(t_\gamma) \big\rvert^2 \lvert \Hel^{t,t} \rvert^2 + \big\lvert P_V(u_\gamma) \big\rvert^2 \lvert \Hel^{u,u} \rvert^2 \notag \\
    & + 2 \Re\big[ P_V(t_\gamma) P_V^*(u_\gamma) \big] \lvert \Hel^{t,u} \rvert^2 \Big) \notag \\
    + & \sum_{\{V_1,V_2\}} 2C_{V_1} C_{V_2} \Big( \Re\big[ P_{V_1}(t_\gamma) P_{V_2}^*(t_\gamma) \big] \lvert \Hel^{t,t} \rvert^2 \notag \\
    & + \Re\big[ P_{V_1}(u_\gamma) P_{V_2}^*(u_\gamma) \big] \lvert 
    \Hel^{u,u} \rvert^2  \\
    & + \Re\big[ P_{V_1}(t_\gamma) P_{V_2}^*(u_\gamma) + P_{V_1}(u_\gamma) P_{V_2}^*(t_\gamma) \big] \lvert \Hel^{t,u} \rvert^2 \Big) \bigg], \notag
\end{align}
where the second sum extends over $\{V_1,V_2\} = \{\rho,\omega\}, \{\rho,\phi\}, \{\omega,\phi\}$ and we introduced
\begin{align}
    \lvert \Hel^{t,t} \rvert^2 
    &= g^{\alpha_1 \tilde{\alpha}_1} g^{\alpha_2 \tilde{\alpha}_2} \Hel_{\alpha_1 \alpha_2}^{t} \Hel_{\tilde{\alpha}_1 \tilde{\alpha}_2}^{t}, \notag \\
    \lvert \Hel^{u,u} \rvert^2 
    &= g^{\alpha_1 \tilde{\alpha}_1} g^{\alpha_2 \tilde{\alpha}_2} \Hel_{\alpha_1 \alpha_2}^{u} \Hel_{\tilde{\alpha}_1 \tilde{\alpha}_2}^{u}, \notag \\
    \lvert \Hel^{t,u} \rvert^2 
    &= g^{\alpha_1 \tilde{\alpha}_2} g^{\alpha_2 \tilde{\alpha}_1} \Hel_{\alpha_1 \alpha_2}^{t} \Hel_{\tilde{\alpha}_1 \tilde{\alpha}_2}^{u}.
\end{align}
As in \autoref{eq:form_factor_DP_VW}, the propagators $P_V(x)$ are to be understood as BW propagators for all $V$ in the CW approximation and BW propagators for $V = \omega, \phi$ but dispersively improved variants for $V = \rho$ in the variant VW.
Inserting the kinematics of the process, these expressions simplify to
\begin{align}
    \lvert \Hel^{t,t} \rvert^2
    &= \lvert \Hel^0 \rvert^2 + t_\gamma^2 (s^2 + u_\gamma^2), \notag \\
    \lvert \Hel^{u,u} \rvert^2
    &= \lvert \Hel^0 \rvert^2 + u_\gamma^2 (s^2 + t_\gamma^2), \notag \\
    \lvert \Hel^{t,u} \rvert^2
    &= \lvert \Hel^0 \rvert^2 + t_\gamma u_\gamma (s^2 + t_\gamma u_\gamma),
\end{align}
where we defined
\begin{align}
    \lvert \Hel^0 \rvert^2 
    &= \massOut^4 \big(s^2 + t_\gamma^2 + u_\gamma^2 + 2 s t_\gamma + 2 s u_\gamma + 4 t_\gamma u_\gamma \big) \notag \\
    &\,\,\,\, - 2\massOut^2 \Sigma_\gamma t_\gamma u_\gamma - 2 \massOut^6 \Sigma_\gamma + \massOut^8.
\end{align}

Finally, we consider the normalized semileptonic branching ratios
\beq\label{eq:BR_normalized}
    \widehat{\BR}(\eta^{(\prime)} \to [\pi^0/\eta] \ell^+ \ell^-)
    = \frac{\BR(\eta^{(\prime)} \to [\pi^0/\eta] \ell^+ \ell^-)}{\BR(\eta^{(\prime)} \to [\pi^0/\eta] \gamma \gamma)},
\eeq
which are particularly useful from the theoretical point of view, since they reduce the effect of the uncertainties from the coupling constants.

We perform the phase-space integrations of the differential decay widths, \autoref{eq:DW} and \autoref{eq:DW_gamma_gamma}, numerically with the \Lat{Cuhre} and \Lat{Vegas} algorithm from the \Lat{Cuba} library~\cite{Hahn:2004fe}.
For the numerical evaluation of the PV functions contained in the quantities $\M_V^{u(0)v}$, see \autoref{eq:PV_result_amplitude}, we use \Lat{Collier}~\cite{Denner:2002ii,Denner:2005nn,Denner:2010tr,Denner:2016kdg}.\footnote{%
    A C\texttt{++} interface to the native Fortran library \Lat{Collier} written for this purpose, including an executable demo file, is attached as Supplemental Material to this article~\cite{SupplementalMaterialBoth}.}
The integration is carried out following the decomposition of Eqs.~\eqref{eq:PV_result_amplitude_squared} and \eqref{eq:amplitude_squared_gamma_gamma},
\begin{align}\label{eq:DW_sum}
    \Gamma_{(\gamma)} 
    &= C_\rho^2 \Gamma_{\rho,\rho}^{(\gamma)} + C_\omega^2 \Gamma_{\omega,\omega}^{(\gamma)} + C_\phi^2 \Gamma_{\phi,\phi}^{(\gamma)} \notag \\
    &\quad + C_{\rho} C_{\omega} \Gamma_{\rho,\omega}^{(\gamma)} + C_{\rho} C_{\phi} \Gamma_{\rho,\phi}^{(\gamma)} + C_{\omega} C_{\phi} \Gamma_{\omega,\phi}^{(\gamma)}.
\end{align}
Numerical results for the auxiliary quantities $\Gamma_{V_1,V_2}^{(\gamma)}$ are listed in Appendix~\ref{appx:intermediate_results}.\footnote{%
    Using \Lat{LoopTools}~\cite{Hahn:1998yk} for the evaluation of the PV functions, we observed severe numerical instabilities for some integrations in the variant VW.
    These issues were most extreme in $\Gamma_{V_1,V_2}$ with at least one $V_i = \phi$ for the decays $\eta^{(\prime)} \to \pi^0 e^+ e^-$ but also notably problematic in $\Gamma_{\omega,\omega}$ for $\eta' \to \pi^0 e^+ e^-$.
    They can be traced back to problems with the evaluation in certain regions of the phase space and might be related to vanishing \LN{Gram} determinants in the PV reduction procedure, but their exact origin remains obscure to us, in particular because a decomposition into coefficient functions does not improve this behavior and the evaluation with \Lat{Collier} using scalar functions does not suffer from such instabilities.}
    
%---------------------------------------------------------------------------------------------------
\section{Scalar rescattering contributions}
\label{sec:rescattering}
%---------------------------------------------------------------------------------------------------
While there are good reasons to assume that the VMD model captures the most significant contributions to the semileptonic $\eta^{(\prime)}$ decays, we will assess scalar rescattering contributions explicitly by calculating them for the $\eta \to \pi^0 \ell^+ \ell^-$ channels.
For the $\eta'$ channels, the vector mesons have sufficient energy to go quasi on-shell, so that an even stronger dominance of the VMD mechanism is expected.

%---------------------------------------------------------------------------------------------------
\subsection{Isolating the \boldmath{$S$}-wave in the hadronic sub-amplitude}
%---------------------------------------------------------------------------------------------------
With the decay $\eta \to \pi^0 \ell^+ \ell^-$ being driven by the two-photon intermediate state, as discussed in \autoref{sec:intro}, the hadronic sub-process we consider is again $\eta \to \pi^0 \gamma \gamma$.
The corresponding sub-amplitude $\Hel_{\lambda\lambda'}$, defined in \autoref{eq:def_hel_amp}, can be expressed in terms of the tensor amplitude $\Hel_{\mu\nu}$ according to
\begin{align}
    e^{\iu(\lambda - \lambda') \varphi} \Hel_{\lambda \lambda'}
    &= \eps_{\lambda}^{*\mu}(q_1) \eps_{\lambda'}^{*\nu}(q_2) \, \Hel_{\mu \nu}.
\end{align}
In the following, we choose
\begin{align}\label{eq:pol_vectors}
    \eps^\mu_\pm(q_1)
    &= \frac{1}{\sqrt{2}} (0,\mp 1,-\iu,0), \notag \\
    \quad \eps^\mu_\pm(q_2)
    &= \frac{1}{\sqrt{2}} (0,\mp 1,\iu,0)
\end{align}
as the explicit form for the polarization vectors.
In the context of the hadronic process $\eta \to \pi^0 \gamma \gamma$, we use the \LN{Mandelstam} variables $s$, $t_\gamma$, and $u_\gamma$ as defined in \autoref{eq:Mandelstam_gamma}.
For on-shell photons, the tensor amplitude $\Hel^{\mu\nu}$ can be written in terms of two independent tensor structures $T_{1/2}^{\mu\nu}$~\cite{Lu:2020qeo}, 
\begin{align}\label{eq:tensor_structures_T1_T2_pi_eta_to_gamma_gamma}
    T_1^{\mu\nu}
    &= \frac{1}{2} s \, g^{\mu\nu} - q_2^\mu q_1^{\nu}, \notag \\
    T_2^{\mu\nu}
    &= 2s \Delta^\mu \Delta^\nu + 4(q_1 \Delta)(q_2 \Delta) g^{\mu\nu} \notag \\
    &\quad -4(q_2 \Delta) \Delta^\mu q_1^\nu -4(q_1 \Delta) q_2^\mu \Delta^\nu,
\end{align}
with $\Delta^\mu = (P + p_0)^\mu$, which manifestly fulfill the necessary \LN{Ward} identities.
The expansion of the tensor amplitude in this basis involves two scalar amplitudes $A$ and $B$ and reads
\begin{align}\label{eq:tensor_amp_pi_eta_to_gamma_gamma}
    \Hel^{\mu\nu} 
    &= A(s,t_\gamma) T_1^{\mu\nu} + B(s,t_\gamma) T_2^{\mu\nu}.
\end{align}
Contracting the tensor amplitude \eqref{eq:tensor_amp_pi_eta_to_gamma_gamma} with the polarization vectors gives an expression for the helicity amplitudes in terms of the scalar amplitudes, 
\begin{align}\label{eq:hel_amp_scalar_amp_pi_eta_to_gamma_gamma}
    \Hel_{++}(s,t_\gamma)
    &= -\frac{s}{2} A(s,t_\gamma) - s \big[ 2(\Meta^2 + \! \MpiN^2) - s \big] B(s,t_\gamma), \notag \\
    \Hel_{+-}(s,t_\gamma)
    &= \big[ (t_\gamma-u_\gamma)^2 - \lambda_{\pi^0 \eta}(s) \big] B(s,t_\gamma).
\end{align}
Here and in the following, we use the abbreviation 
\begin{align}
    \lambda_{P_1P_2}(s) 
    &\equiv \lambda(s,M_{P_1}^2,M_{P_2}^2).
\end{align}
To isolate the $S$-wave, we will neglect $D$- and higher partial waves, including the whole $\Hel_{+-}$ contribution, since its partial-wave expansion starts with $D$-waves. 
Consequently, we are required to set the scalar amplitude $B$ to zero, which leads to the $S$-wave contributing only through the tensor structure $T_1^{\mu\nu}$.
Furthermore, setting $B$ to zero allows us to use the $S$-wave amplitude $h^{L=0}_{++}$ to fix the scalar amplitude $A$ via \autoref{eq:hel_amp_scalar_amp_pi_eta_to_gamma_gamma},
\begin{align}\label{eq:scalar_S_wave_amp_pi_eta_to_gamma_gamma}
    A^0(s)
    &= -\frac{2}{s} \, h_{++}^0(s).
\end{align}
Note that the $(++)$ helicity amplitude has a soft-photon zero at $s=0$, such that $A^0(s)$ has no singularity at that point despite the factor $1/s$.
\begin{figure}[t]
	\centering
	\includegraphics[scale=0.95]{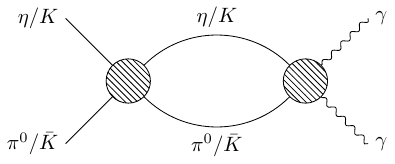}
	\caption{The two intermediate states $\pi^0\eta/K\bar{K}$ contributing to the two-photon amplitudes.
    The dispersive representation of those amplitudes is constructed in Ref.~\cite{Lu:2020qeo}.}
	\label{fig:intermediate_states}
\end{figure}

%---------------------------------------------------------------------------------------------------
\subsection{Rescattering effects in the hadronic sub-process}
%---------------------------------------------------------------------------------------------------

In Ref.~\cite{Lu:2020qeo}, the rescattering effects in $\eta \to \pi^0 \gamma \gamma$ are described by means of a coupled-channel analysis, taking into account $\pi^0\eta$ and $K \bar{K}$ intermediate states; \Lat{cf.}\ Fig.~\ref{fig:intermediate_states}.
Using the \LN{Omn\`es} matrix $\Omega(s)$ for the $\pi^0\eta/(K\bar{K})_{I=1}$ system constructed therein, one can write a dispersive representation for the $S$-wave amplitudes,
\begin{widetext}
\begin{align}\label{eq:dispersive_rep_of_h0++}
    \binom{h_{++}^0(s)}{k^0_{1,++}(s)} 
    &= \Omega(s) \Bigg\{ \binom{a}{b} \, s
    +
    \frac{s^2}{\pi} \Bigg( \sum_V \int\limits_{-\infty}^{s_V} \! \dff z \, \frac{\Omega^{-1}(z)}{z^2 (z-s)} \, \Im \binom{h^{0,V}_{++}(z)}{k^{0,V}_{++}(z)}
    -
    \int\limits_{s_{\pi\eta}}^{\infty} \! \dff z \, \frac{\Im\big( \Omega^{-1}(z) \big)}{z^2 (z-s)} \binom{0}{k^{0,\Born}_{1,++}(z)} \Bigg) \Bigg\},
\end{align}
\end{widetext}
with $s_{\pi\eta} = (\Meta + \MpiN)^2$ the threshold for the $\pi^0\eta$ intermediate state and
\begin{align}
    s_V
    = -\frac{1}{M_V^2} \big( M_V^2 - \Meta^2 \big) \big( M_V^2 - \MpiN^2 \big)
\end{align}
the onset of the left-hand cut.
Here, we include the VMD contributions from the $\rho$, $\omega$, and $\phi$ mesons for the $\pi^0\eta$ channel ($h^{0,V}_{++}$) and the $K^*$ for the $K\bar{K}$ channel ($k^{0,V}_{++}$) in the zero-width approximation.
Using the polarization vectors \eqref{eq:pol_vectors} and the coupling constants $C_V$ defined in \autoref{eq:PV_result_amplitude}, the VMD amplitude for the $\pi^0\eta$ channel for photons with polarization $(++)$ as well as the corresponding $S$-wave amplitude are given by
\begin{align}\label{eq:hel_amp_++}
    \Hel_{++}^V (s,t_\gamma)
    &= \frac{C_V}{4} \frac{st_\gamma}{M_V^2 - t_\gamma - \iu \eps} + (t_\gamma \leftrightarrow u_\gamma), \notag \\
    h_{++}^{0,V}(s)
    &= \frac{C_V}{2} \bigg( \frac{s M_V^2}{\lambda^{1/2}_{\pi^0 \eta}(s)} \log\bigg[ \frac{X_V(s) + 1}{X_V(s) - 1} \bigg] - s \bigg) , \notag \\
    X_V(s)
    &= \frac{2M_V^2 - (\Meta^2 + \MpiN^2) + s}{\lambda_{\pi^0\eta}^{1/2}(s)} \, .
\end{align}
The logarithm in \autoref{eq:hel_amp_++} induces the left-hand cut starting from $s_V$.
The VMD contribution to the $K\bar{K}$ channel, $K_{++}^{V}$, can be treated in complete analogy.
In the $K\bar{K}$ channel, the QED \LN{Born} term projected onto isospin $I=1$ is included in addition,
\begin{align}
    K_{1,++}^{\Born}(s,t_\gamma)
    &= \frac{\sqrt{2} \, s \, \MK^2}{(t_\gamma - \MK^2)(u_\gamma - \MK^2)}, \notag \\
    k_{1,++}^{0,\Born}(s)
    &= \frac{2 \sqrt{2} \MK^2}{s \, \sigma_K(s)}
    \log\bigg[ \frac{1 + \sigma_K(s)}{1 - \sigma_K(s)} \bigg],
\end{align}
with $\sigma_{K}(s) \equiv \sqrt{1 - 4 \MK^2/s}$.
In \autoref{eq:dispersive_rep_of_h0++}, the soft-photon zero is already taken care of; the remaining subtraction constants $a$ and $b$ are determined in accordance with Ref.~\cite{Lu:2020qeo}, where an \LN{Adler} zero at $s_A = \Meta^2$ is implemented to fix one of these and the other one is fit to experimental data.

Subtracting the VMD contributions \eqref{eq:hel_amp_++} from the complete $S$-wave amplitude $h^0_{++}$ \eqref{eq:dispersive_rep_of_h0++} allows us to isolate the rescattering effects in \autoref{eq:scalar_S_wave_amp_pi_eta_to_gamma_gamma},
\begin{align}
    A^0_{\resc}(s)
    &= -\frac{2}{s} \, \bigg( h_{++}^0(s) -\!\!\! \sum_{V = \rho,\omega,\phi} \!\!\! h^{0,V}_{++}(s) \bigg).
\end{align}
With this, we can now construct the $S$-wave tensor amplitude containing only the rescattering contributions,
\begin{align}\label{eq:rescattering_tensor_amplitude_S_wave_pi_eta_to_gamma_gamma}
    \widetilde{\Hel}^{\mu\nu}
    &= A^0_{\resc}(s) \, T_1^{\mu\nu}.
\end{align}

%---------------------------------------------------------------------------------------------------
\subsection{Loop calculation}
%---------------------------------------------------------------------------------------------------
\begin{figure}[t]
	\centering
	\includegraphics[scale=0.95]{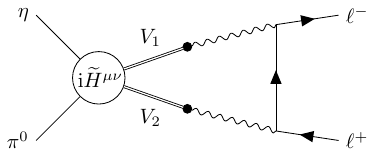}
	\caption{The triangle loop contributing to $\pi^0 \eta \to \ell^+ \ell^-$, which contains the tensor amplitude $\widetilde{H}^{\mu \nu}$ that captures the rescattering effects in $\pi^0 \eta \to \gamma \gamma$, with the photon virtualities modeled via vector-meson propagators.
    This process is related to the corresponding $\eta$ decay via crossing symmetry.}
	\label{fig:rescattering_loop}
\end{figure}
In order to calculate the contribution of $S$-wave rescattering effects to the decay $\eta \to \pi^0 \ell^+\ell^-$, we retain the tensor amplitude \eqref{eq:rescattering_tensor_amplitude_S_wave_pi_eta_to_gamma_gamma} for the $\eta \to \pi^0 \gamma \gamma$ vertex.
This reduces the loop from a box to a triangle topology; see \autoref{fig:rescattering_loop}.
We denote the tensor QED sub-amplitude for $\gamma \gamma \to \ell^+ \ell^-$ by $L^{\mu\nu}$.
At tree level, the construction is straightforward, and after simplifying with \autoref{eq:Dirac_eq}, one finds
\begin{align}\label{eq:QED_tensor_amp_gamma_gamma_to_ll}
    L^{\mu\nu}
    &= -\bar{u}_s \frac{2p_-^\mu - \gamma^\mu \slashed{q_1}}{(p_- - q_1)^2 - m_\ell^2 + \iu \eps} \gamma^\nu v_r.
\end{align} 
Note that we do not have to concern ourselves with calculating the $S$-wave projection of the QED sub-amplitude $\gamma^* \gamma^* \to \ell^+ \ell^-$, since the loop integration will take care of the projection automatically.
Furthermore, to avoid double counting, we do not include the crossed channel, which is described by the same amplitude due to the symmetry of the triangle loop.

When taking into account the photon virtualities, the gauge-invariant tensor structure $T_1^{\mu\nu}$, in particular, acquires additional terms~\cite{Bardeen:1968ebo,Tarrach:1975tu,Colangelo:2015ama}, 
\begin{align}
    T_1^{\mu\nu}(q_1^2,q_2^2)
    &= \frac{1}{2}\big( s - q_1^2 - q_2^2 \big) g^{\mu\nu} - q_2^\mu q_1^\nu.
\end{align}
The impact of the photon virtualities is then further modeled by including factors $M_V^2 P_V^\BW(q^2)$ for both photons, resulting in a hadronic tensor amplitude for off-shell photons on the basis of the on-shell one,
\begin{align}\label{eq:photon_virtualities_pi_eta_to_gamma_gamma}
    \widetilde{\Hel}^{\mu\nu}(q_1^2,q_2^2)
    &= M_{V_1}^2 P_{V_1}^\BW(q_1^2) M_{V_2}^2 P_{V_2}^\BW(q_2^2) \notag \\
    &\quad \times A_{\resc}^0(s)T_1^{\mu\nu}(q_1^2,q_2^2).
\end{align}
This is a naive generalization to virtual photons that corresponds to a scalar-resonance approximation.
It avoids the known complications, \Lat{e.g.}, from the modified partial-wave projections of the VMD amplitudes; see Refs.~\cite{Moussallam:2013una, Moussallam:2021dpk} for a more rigorous treatment.
We deem this approximation sufficient in the context of the semileptonic decays.
The prescription in \autoref{eq:photon_virtualities_pi_eta_to_gamma_gamma} is consistent with the monopole model for the form factors constructed in \autoref{sec:form_factors}.

The rescattering contribution to the $\eta \to \pi^0 \ell^+ \ell^-$ amplitude is then given by 
\begin{align}
    \iu \, \widetilde{\M}(s)
    &= \Big(\frac{\alpha}{\pi}\Big)^2 \int \dff^4 q_1 \, \frac{\widetilde{\Hel}^{\mu\nu}(q_1^2,q_2^2)}{q_1^2 + \iu \eps}\frac{L_{\mu\nu}}{q_2^2 + \iu \eps},
\end{align}
with $q_2 = p_+ + p_- - q_1$. 

Understanding the $S$-wave amplitude as an enhancement due to the $a_0(980)$ resonance with $I^G(J^{PC}) = 1^- (0^{++})$, only the combination of $\rho$ and $\omega$ is allowed for the vector mesons $V_1$ and $V_2$.
With that, the $S$-wave rescattering contribution is given by
\begin{align}
    \widetilde{\M}(s)
    &= -\iu\, \Big(\frac{\alpha}{\pi}\Big)^2 \Mrho^2 \Momega^2 A^0_{\mathrm{resc}}(s) \\
    &\quad \times \int \dff^4 q_1 \, P_{\rho}^\BW(q_1^2) \, P_{\omega}^\BW(q_2^2) \, \frac{T_1^{\mu\nu}(q_1^2,q_2^2) \, L_{\mu\nu}}{(q_1^2 + \iu \eps)(q_2^2 + \iu \eps)}. \notag
\end{align}
Note that with $T_1^{\mu\nu} \propto \Order(q_1^2)$, the integral is convergent only due to the dependence on the photon virtualities introduced in \autoref{eq:photon_virtualities_pi_eta_to_gamma_gamma}.
This is a consequence of the reduction from a box to a triangle loop.
Contracting the tensor structures and performing a PV decomposition allows us to separate a factor of $m_{\ell} s/(\Mrho^2 \Momega^2)$ with only the $\bar{u}_s v_r$ spinor structure from \autoref{eq:PV_result_amplitude} contributing,
\begin{align}\label{eq:rescattering_contibution_semileptonic_amplitude}
    \widetilde{\M}(s) 
    &= \iu \, (4\pi\alpha)^2 s\, A^0_{\resc}(s) \widetilde{\M}_\had^{uv}(s) \M_\QED^{uv}.
\end{align}
Here, $\widetilde{\M}_\had^{uv}(s)$ contains the remaining PV master integrals.

%---------------------------------------------------------------------------------------------------
\section{Results and discussion}
\label{sec:results_discussion}
%---------------------------------------------------------------------------------------------------
We present the results for the semileptonic decays in the form of branching ratios as well as singly- and doubly-differential decay widths.
The branching ratios are particularly apt to demonstrate the effects of the different form-factor models.
Furthermore, we examine the contribution of scalar rescattering effects to the branching ratios and normalize these to the corresponding two-photon analogs.
For all of our results, the quoted uncertainties stem from the experimental uncertainties that enter via the coupling constants and amount to $\sim 10 \perc$.
The uncertainties from the numerical integration, on the other hand, are at least one order of magnitude smaller and therefore omitted. 

%----------------------------------------------------------------------------------------------
\subsection{Differential decay widths}
\label{sec:discussion_diff}
%----------------------------------------------------------------------------------------------
\begin{figure*}[t]
	\centering
	\includegraphics[width=0.75\textwidth]{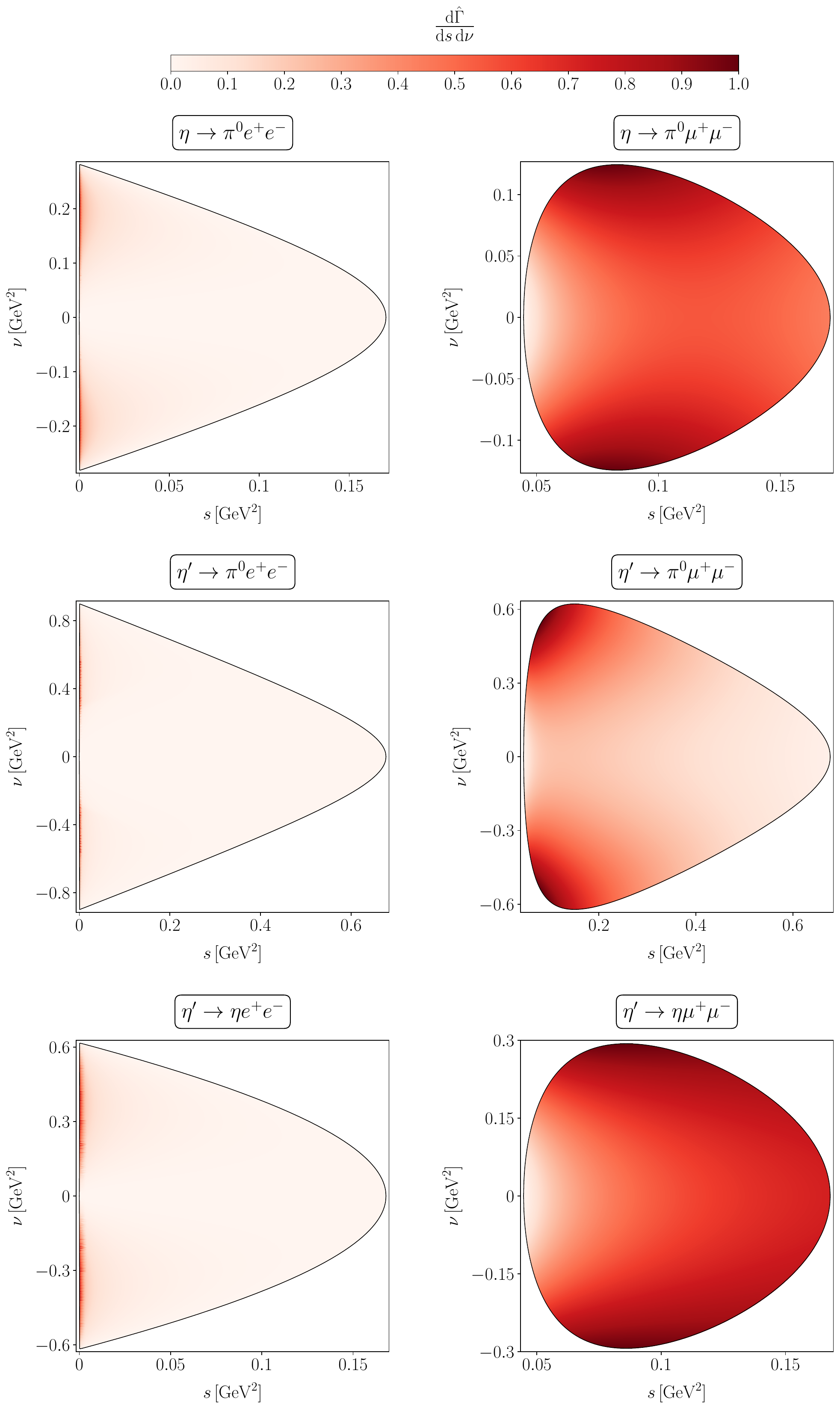}
	\caption{\LN{Dalitz} plots for the MP model in the variant CW, normalized to the maximum value within the available phase space of the respective channel, $\dff \hat{\Gamma}/\dff s \, \dff \nu = [\dff \Gamma/\dff s \, \dff \nu]/[\max\, \dff \Gamma/\dff s \, \dff \nu]$.}
	\label{fig:dalitz}
\end{figure*}
\begin{figure*}[t]
	\centering
	\includegraphics[width=\textwidth]{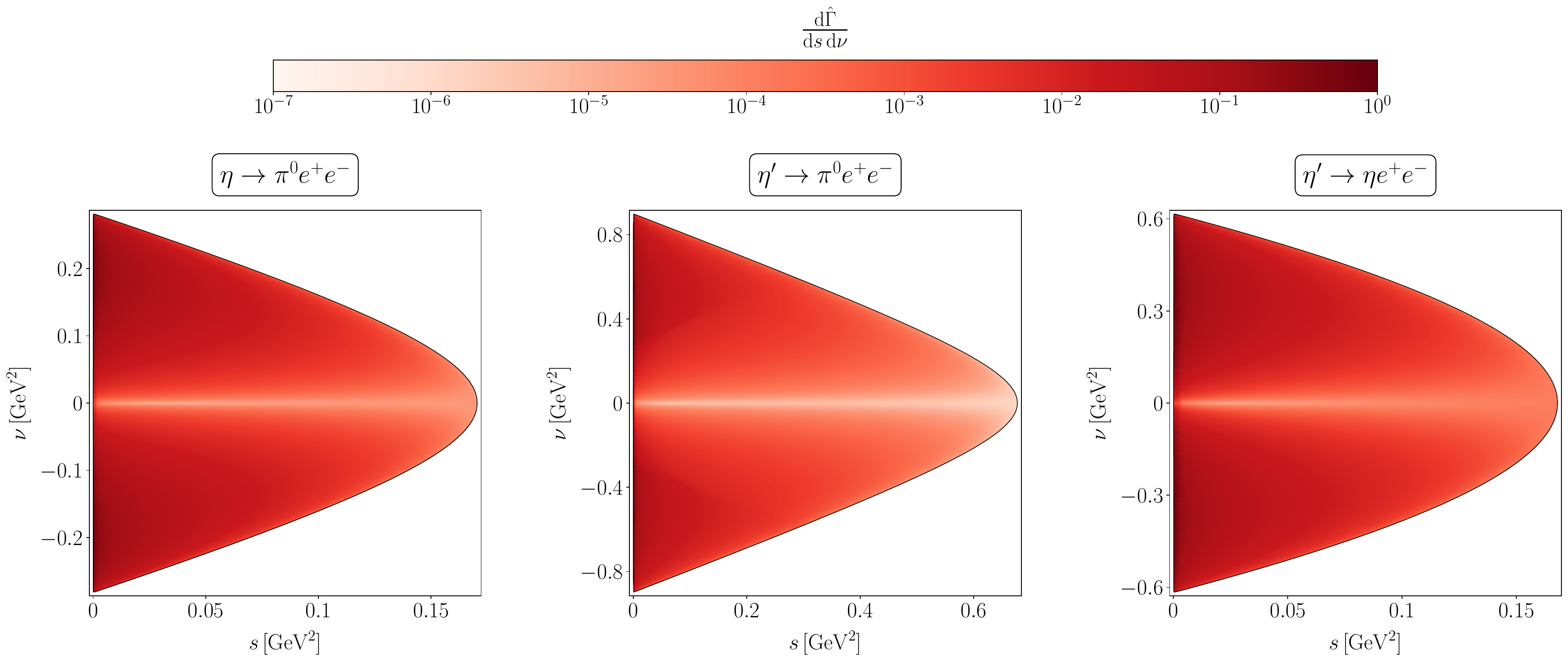}
	\caption{Logarithmic \LN{Dalitz} plots for the electron channels with the MP model in the variant CW, normalized to the respective maximum value within the available phase space; see \autoref{fig:dalitz}.}
	\label{fig:dalitz_log}
\end{figure*}
\begin{figure*}[t]
	\centering
	\includegraphics[width=0.95\textwidth]{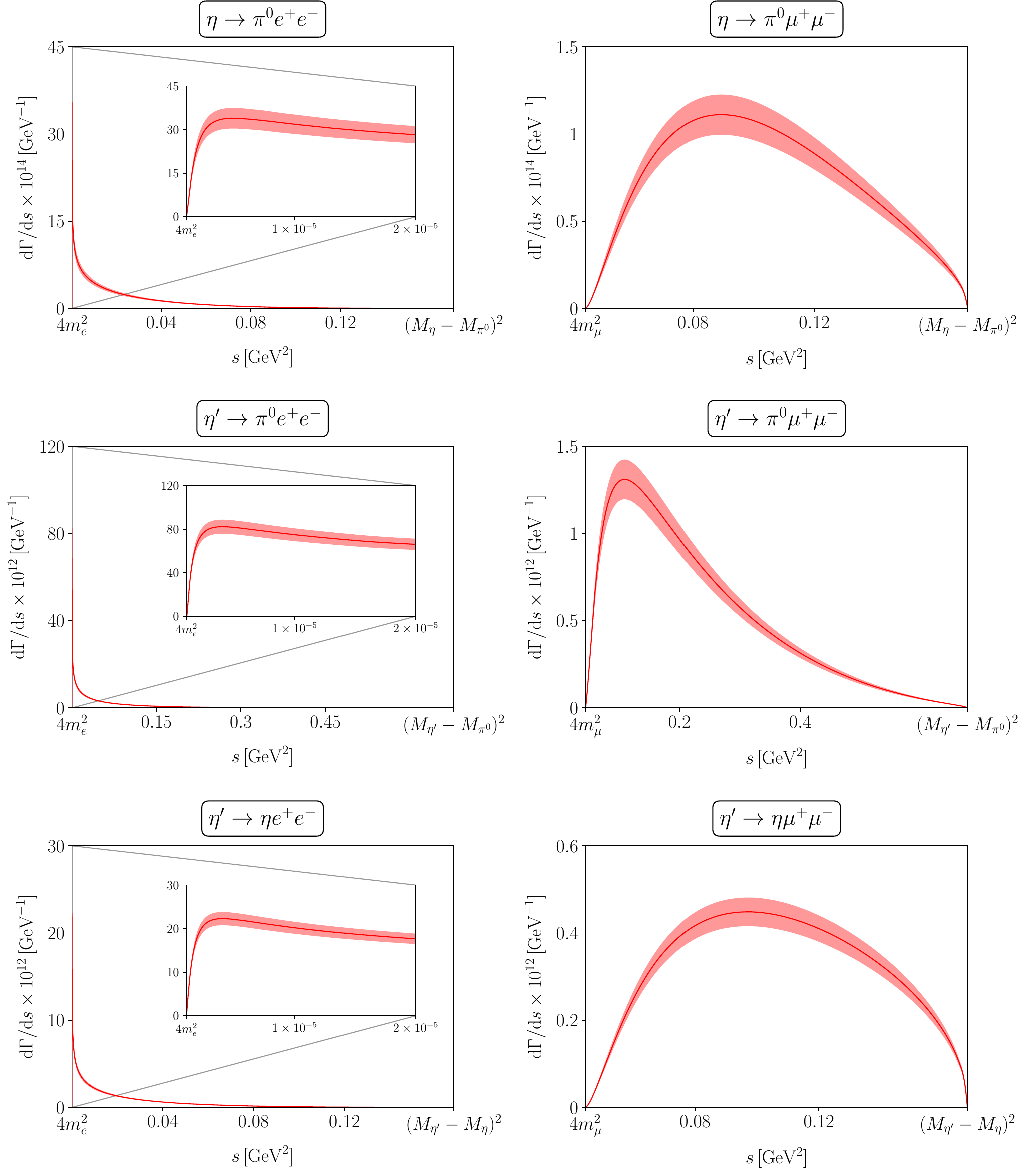}
	\caption{Singly-differential decay widths in the \LN{Mandelstam} variable $s$, obtained with the MP model in the variant CW.
    Here, the inlays amplify the behavior close to the lower threshold of the phase space, where the distribution shows a strong peak for the channels with electrons in the final state.
    The uncertainty is entirely due to the dominant phenomenological uncertainty of $\lvert \F_{V P}(0) \rvert$; see \autoref{tab:couplings_pheno}.}
	\label{fig:diff_DW}
\end{figure*}
The doubly- and singly-differential distributions of the semileptonic decays exhibit distinct characteristics, with the most prominent differences being observable between the decays with electrons and muons in the final state; see Figs.~\ref{fig:dalitz}--\ref{fig:diff_DW}.
While the majority of the doubly-differential distribution for the electron channels is contained in a small fraction close to the threshold in the invariant lepton mass, the decays with muons in the final state display a spread-out distribution that covers large parts of the available phase space.
For the electron final state, in particular, it is important to take account of the region close to the threshold in the invariant lepton mass both when integrating over the phase space and when performing a measurement, as significant parts of the decay width are readily missed otherwise.
Furthermore, the logarithmic scale shows that the distributions possess a minimum for $\nu = 0$, where $\nu \propto \cos\theta_s$, with $\theta_s$ the $s$-channel scattering angle.
With only even partial waves contributing to the decays, this feature can be attributed to the dominance of $D$-waves over the helicity-suppressed $S$-waves---which do not show such an angular distribution---whereas for the muon channels, this suppression is less pronounced.
Beyond the difference in the final-state leptons, the principal visible differentiations concern the size of the phase space, which is significantly larger for $\eta' \to \pi^0 \ell^+ \ell^-$ than for $\eta \to \pi^0 \ell^+ \ell^-$ and $\eta' \to \eta \ell^+ \ell^-$.

For all decay channels, the obtained \LN{Dalitz} plots do not follow a flat distribution, which was assumed for the experimental analysis of $\eta \to \pi^0 e^+ e^-$ in Ref.~\cite{WASA-at-COSY:2018jdv}.
This assumption is justified for a potential $\CC$-violating contribution~\cite{Akdag:2023pwx} but inaccurate for the standard-model result; we therefore propose a reevaluation of the experimental data and a reassessment of the reported upper limit.

The singly-differential distributions for the electron channels explicitly resolve a strongly peaked structure for invariant lepton masses close to the threshold and a subsequent rapid decrease.
For muons in the final state, the singly-differential distribution is much different, with a broad peak that is situated more centrally in the phase space.
This behavior is in correspondence with the observation that for $m_\ell \approx 0$, the threshold in $s$ approximately collapses to the threshold of the two-photon intermediate state, $s = 0$, where the two-photon cut induces a behavior $\propto \log(s)$~\cite{Cheng:1967zza}.
Hence, for the electron final state, this logarithmic divergence manifests itself as a peak close to the threshold in $s$, regularized by a phase-space factor and forced to zero at $s = 4m_\ell^2$, see \autoref{eq:Mandelstam_limits}, whereas the muon channels have a much higher threshold, far from the logarithmic divergence.

%----------------------------------------------------------------------------------------------
\subsection{Branching ratios in the different models}
\label{sec:discussion_BRs}
%----------------------------------------------------------------------------------------------
The sensitivity of the semileptonic decays to the different form-factor parameterizations, \Lat{i.e.}, a point-like, monopole, or dipole interaction, each with constant or energy-dependent widths, can be probed by comparing the results for the branching ratios collected in \autoref{tab:BRs}.

Our results for the decays $\eta \to \pi^0 \ell^+ \ell^-$ obtained with constant form factors and widths are compatible with the results of Ref.~\cite{Escribano:2020rfs}, which similarly assumed a point-like interaction.
Instead of determining the coupling constants purely from phenomenology, the authors modeled these using a symmetry-driven quark model, which results in only slightly different numerical values. 
For the $\eta'$ decays, on the other hand, we find significant disagreement, which might be due to numerical difficulties when calculating the box diagrams in a non-automated way via \LN{Feynman} parameters.
\begin{table}[t]
	\centering
    {
    \setlength{\tabcolsep}{1.5pt}
	\begin{tabular}{l  p{0.7cm}  r  c  r  c  r  c  r}
	\toprule
		& & \multicolumn{7}{c}{Branching ratio$/10^{-9}$} \\
        & & \multicolumn{1}{c}{PL} & & \multicolumn{1}{c}{MP} & & \multicolumn{1}{c}{DP} & & \multicolumn{1}{c}{\!\!\!Ref.~\cite{Escribano:2020rfs}} \\
		\midrule
        \multirow{2}{*}{$\eta \to \pi^0 e^+ e^-$} & CW & 
        $2.10(23)$ 
        & &
        $1.35(15)$ 
        & & 
        $1.33(15)$ 
        & &
        %\multirow{2}{*}{$2.0(1)(1)(1)$} \\ % KEEP!
        \multirow{2}{*}{$2.0(2)$} \\
        & VW &
        $2.06(22)$
        & &
        $1.40(15)$ 
        & &
        $1.36(15)$ 
        & & 
        \\
        \\[-0.25cm]
        \multirow{2}{*}{$\eta \to \pi^0 \mu^+ \mu^-$} & CW &
        $1.37(15)$
        & &
        $0.70(8)$ 
        & &
        $0.66(7)$ 
        & &
        %\multirow{2}{*}{$1.1(1)(1)(1)$} \\ % KEEP!
        \multirow{2}{*}{$1.1(2)$} \\
        & VW &
        $1.32(14)$
        & &
        $0.71(8)$ 
        & & 
        $0.67(7)$ 
        & & 
        \\
        \midrule
        \multirow{2}{*}{$\eta' \to \pi^0 e^+ e^-$} & CW &
        $3.82(33)$
        & &
        $3.08(27)$ 
        & &
        $3.14(27)$ 
        & & 
        %\multirow{2}{*}{$4.5(3)(4)(4)$} \\ % KEEP!
        \multirow{2}{*}{$4.5(6)$} \\
        & VW &
        $3.81(33)$
        & &
        $3.30(28)$
        & & 
        $3.30(28)$ 
        & & 
        \\
        \\[-0.25cm]
        \multirow{2}{*}{$\eta' \to \pi^0 \mu^+ \mu^-$} & CW &
        $2.57(23)$
        & &
        $1.69(15)$ 
        & &
        $1.68(15)$ 
        & &
        %\multirow{2}{*}{$1.7(1)(2)(2)$} \\ % KEEP!
        \multirow{2}{*}{$1.7(3)$} \\
        & VW &
        $2.53(23)$
        & &
        $1.81(16)$
        & &
        $1.81(16)$ 
        & & 
        \\
        \midrule
        \multirow{2}{*}{$\eta' \to \eta e^+ e^-$} & CW &
        $0.53(4)$
        & &
        $0.48(4)$ 
        & &
        $0.49(4)$ 
        & &
        %\multirow{2}{*}{$0.43(3)(2)(18)$} \\ % KEEP!
        \multirow{2}{*}{$0.4(2)$} \\
        & VW &
        $0.51(4)$
        & &
        $0.50(4)$
        & &
        $0.50(4)$ 
        & &
        \\
        \\[-0.2cm]
        \multirow{2}{*}{$\eta' \to \eta \mu^+ \mu^-$} & CW &
        $0.287(26)$
        & &
        $0.213(18)$
        & &
        $0.207(18)$
        & &
        %\multirow{2}{*}{$0.15(1)(1)(5)$} \\ % KEEP!
        \multirow{2}{*}{$0.15(5)$} \\
        & VW &
        $0.280(25)$
        & &
        $0.225(20)$
        & &
        $0.240(21)$
        & &
        \\
		\bottomrule
	\end{tabular}
    }
	\caption{The branching ratios of the semileptonic decays, \autoref{eq:BR}, resulting for the models PL, MP, and DP in both variants CW and VW.
    The uncertainty is entirely due to the dominant experimental uncertainty of $\lvert \F_{V P}(0) \rvert$; see \autoref{tab:couplings_pheno}.
    For reference, we also give the corresponding results from Ref.~\cite{Escribano:2020rfs}, where we added the quoted uncertainties in quadrature.}
	\label{tab:BRs}
\end{table}

Implementing non-trivial form factors leads to a significant decrease of the branching ratio for all decays, with the muon channels being subject to a larger reduction than the electron channels and the $\eta'$ decays to less reduction than the $\eta$ decays.
More specifically, the decrease amounts to $\sim 35\perc$ for $\eta \to \pi^0 e^+ e^-$ and $\sim 50\perc$ for $\eta \to \pi^0 \mu^+ \mu^-$. 
For $\eta' \to \pi^0 \ell^+ \ell^-$, the branching ratios are reduced by $\sim 20\perc$ for electrons and $\sim 35\perc$ for muons in the final state.
Regarding $\eta' \to \eta \ell^+ \ell^-$, the branching ratios decrease by $\sim 10\perc$ for electrons and $\sim 25\perc$ for muons in the final state.
This gives strong indication that the photon virtualities cannot be neglected in the analyzed processes, since constant form factors are likely to overestimate the decay widths.

The dipole form factors, which feature the expected high-energy behavior $\sim 1/q^4$, further assess the sensitivity on the precise parameterization of the form factors.
Compared to the variation observed between constant form factors and the monopole parameterization, their effect is, however, negligible, leading to a further decrease for $\eta \to \pi^0 \ell^+ \ell^-$, $\eta' \to \pi^0 \mu^+ \mu^-$, and $\eta' \to \eta \mu^+ \mu^-$ and a slight increase for $\eta' \to \pi^0 e^+ e^-$ and $\eta' \to \eta e^+ e^-$, both at most at the level of $5\perc$.

Using spectral representations to implement energy-dependent widths for the broad vector mesons, \Lat{i.e.}, $\rho^{(\prime)}$, $\omega'$, and $\phi'$, leads to a decrease in the branching ratio of less than $4\perc$ for all decays with constant form factors and an increase of not more than $8\perc$ both in the monopole and dipole models, with the exception of $\eta' \to \eta \mu^+ \mu^-$, where the increase even reaches $\sim 15\perc$.

All these variations are small compared to the difference between the results in the PL model and any other model and mostly even small compared to the phenomenological uncertainties.
We thus infer the semileptonic decays to be rather insensitive to the precise parameterization of the photon virtualities in the form factors.
Therefore, we restricted our discussion of the \LN{Dalitz} and singly-differential plots in \autoref{sec:discussion_diff} to the monopole model, as finer details would not be discernible.

%----------------------------------------------------------------------------------------------
\subsection{Scalar rescattering contributions}
\label{sec:discussion_rescattering}
%----------------------------------------------------------------------------------------------
\begin{table}
{
\setlength{\tabcolsep}{3.5pt}
    \begin{tabular}{l  r  r  r}
	\toprule
	    \multirow{2}{*}{} & \multicolumn{3}{c}{Branching ratio} \\ 
        & \multicolumn{1}{c}{VMD} 
        & \multicolumn{1}{c}{rescattering} & \multicolumn{1}{c}{mixed} \\
		\midrule
        $\eta \to \pi^0 e^+ e^-$  & $1.36(15) \times 10^{-9}$ & $2.5 \times 10^{-13}$ & $4.6 \times 10^{-13}$ \\
        $\eta \to \pi^0 \mu^+ \mu^-$ & $0.67(7) \times 10^{-9}$ & $2.8 \times 10^{-11}$ & $-2.6 \times 10^{-11}$ \\
		\bottomrule
	\end{tabular}
    }
    \caption{The scalar rescattering contributions to the branching ratios of $\eta \to \pi^0 \ell^+ \ell^-$, \autoref{eq:squared_rescattering}, separated into the pure rescattering and mixed term, as well as the corresponding VMD contributions from \autoref{tab:BRs} for comparison.}
 \label{tab:BRs_S_wave}
\end{table}
We have calculated the $S$-wave rescattering contributions exemplarily for the $\eta \to \pi^0 \ell^+ \ell^-$ decay channels. 
Adding these to the VMD amplitude leads to two additional terms on the level of the squared amplitude in the branching ratio: one pure rescattering term and one term mixing rescattering and VMD effects,
\begin{align}\label{eq:squared_rescattering}
    \lvert \M+\widetilde{\M} \rvert^2 
    = \lvert \M \rvert^2 + \lvert \widetilde{\M} \rvert^2 + 2\Re \big( \M \widetilde{\M}^* \big). 
\end{align}
The two contributions to the branching ratios can be found in \autoref{tab:BRs_S_wave}.
For $\eta \to \pi^0 e^+ e^-$, both the rescattering and the mixed contribution are of $\Order(10^{-4})$ compared to the VMD result. 
This seems plausible, given that a spin flip is necessary to couple a scalar resonance to two leptons, resulting in an amplitude proportional to $m_\ell$.
For $\eta \to \pi^0 \mu^+ \mu^-$, the rescattering and mixed contributions are at the level of $5\perc$ in comparison to the VMD contributions, still notably below the uncertainties of the latter. 
In addition, the two contributions have opposite signs, such that they largely cancel, leading to a suppression of $\Order(10^{-3})$. 
In light of the negligible contributions of the rescattering effects, we consider it unnecessary to calculate errors on them.
Apart from the impact of the uncertainties on the coupling constants $C_V$ within the dispersive integral in \autoref{eq:dispersive_rep_of_h0++}, such a calculation would also have to take into account the uncertainties from fixing the subtraction constants as estimated in Ref.~\cite{Lu:2020qeo}.

A similar order of magnitude is expected for the respective corrections to the other decay channels $\eta' \to [\pi^0 / \eta] \ell^+ \ell^-$, an explicit demonstration of which is, however, beyond the scope of this article.

%----------------------------------------------------------------------------------------------
\subsection{Photonic decays and normalized branching ratios}
\label{sec:discussion_photonic}
%----------------------------------------------------------------------------------------------
\begin{table}[t]
	\centering
    {
    \setlength{\tabcolsep}{10.0pt}
	\begin{tabular}{l  r  r}
	\toprule
	    \multirow{2}{*}{} & \multicolumn{2}{c}{Branching ratio$/10^{-4}$} \\
        & \multicolumn{1}{c}{CW} & \multicolumn{1}{c}{VW} \\
		\midrule
        $\eta \to \pi^0 \gamma \gamma$ & $1.21(13)$ & $1.18(13)$ \\
        $\eta' \to \pi^0 \gamma \gamma$ & $27.8(1.7)$ & $28.1(1.8)$ \\
        $\eta' \to \eta \gamma \gamma$ & $1.10(8)$ & $1.10(8)$ \\
		\bottomrule
	\end{tabular}
    }
	\caption{The branching ratios of the two-photon decays, \autoref{eq:BR_gamma_gamma}, in both variants CW and VW.
    The uncertainty is entirely due to the dominant experimental uncertainty of $\lvert \F_{V P}(0) \rvert$; see \autoref{tab:couplings_pheno}.}
	\label{tab:BRs_gamma_gamma}
\end{table}
The primary motivation for calculating the branching ratios for the two-photon decays $\eta^{(\prime)} \to [\pi^0 / \eta] \gamma \gamma$ within our VMD framework is the normalization \eqref{eq:BR_normalized} of the corresponding semileptonic decays. 
Numerical results for these are collected in \autoref{tab:BRs_gamma_gamma} and \autoref{tab:BRs_normalized}, respectively.
Currently, however, there is also thriving interest in resolving a discrepancy arising from an updated experimental measurement of the $\eta \to \pi^0 \gamma \gamma$ decay~\cite{Giovannella:2023}.
The effect of implementing dispersively improved $\rho$ propagators amounts to less than $2\perc$ and is therefore insignificant, as the phenomenological uncertainties range between $(6$--$11)\perc$.

Our branching ratios with constant widths are in agreement with the VMD results of Ref.~\cite{Escribano:2018cwg}; supplementing those with a linear-$\sigma$-model scalar contribution and chiral loops, the authors quote $\BR(\eta \to \pi^0 \gamma \gamma) = 1.35(8) \times 10^{-4}$, $\BR(\eta' \to \pi^0 \gamma \gamma) = 2.91(21) \times 10^{-3}$, and $\BR(\eta' \to \eta \gamma \gamma) = 1.17(8) \times 10^{-4}$ based on empirical couplings.
These results are slightly larger than the plain VMD numbers but still compatible within uncertainties, indicating that the effects of these model extensions are insignificant at the current level of precision~\cite{Escribano:2018cwg}.

The dispersive analysis of $\eta \to \pi^0 \gamma \gamma$~\cite{Lu:2020qeo} referenced in \autoref{sec:rescattering} also includes the $a_2 \equiv a_2(1320)$ tensor resonance as well as isospin-breaking $\pi^+\pi^-$ contributions, with the result $\BR(\eta \to \pi^0 \gamma \gamma) = 1.81\substack{+0.46 \\ -0.33} \times 10^{-4}$ showing a $\sim 50\perc$ discrepancy with the VMD model.
This deviation can be traced back largely to the $a_2$ contribution, suggesting that the impact of this resonance might be relevant for $\eta \to \pi^0 \gamma \gamma$, specifically at very low diphoton invariant masses.

In light of this finding, it is important to note that we have not included any tensor-meson effects for $\eta \to \pi^0 \ell^+ \ell^-$ in \autoref{sec:rescattering}.
For electrons in the final state, the lower threshold in $s$ is close to the two-photon threshold, so that an effect of similar size as in the photonic case is within the bounds of possibility; the higher threshold for muons, on the other hand, is expected to exclude the region where the $a_2$ resonance is most relevant.
For the $\eta'$ decays, the exchanged vector mesons can go quasi on-shell, so that the VMD mechanism is even more likely to dominate the effect of the tensor resonance.
\begin{table}[t]
	\centering
    {
    \setlength{\tabcolsep}{3.5pt}
	\begin{tabular}{l  c  r  c  r  c  r}
	\toprule
		& & \multicolumn{5}{c}{Normalized branching ratio$/10^{-6}$} \\
        & & \multicolumn{1}{c}{PL} & & \multicolumn{1}{c}{MP} & & \multicolumn{1}{c}{DP} \\
		\midrule
        \multirow{2}{*}{$\eta \to \pi^0 e^+ e^-$} & CW & 
        $17.422(28)$ 
        & &
        $11.197(11)$ 
        & & 
        $11.032(9)$ 
        \\
        & VW &
        $17.510(20)$
        & &
        $11.855(7)$ 
        & &
        $11.531(4)$ 
        \\
        \\[-0.25cm]
        \multirow{2}{*}{$\eta \to \pi^0 \mu^+ \mu^-$} & CW &
        $11.371(20)$
        & &
        $5.781(7)$ 
        & &
        $5.450(6)$ 
        \\
        & VW &
        $11.197(25)$
        & &
        $6.020(10)$ 
        & & 
        $5.647(5)$ 
        \\
        \midrule
        \multirow{2}{*}{$\eta' \to \pi^0 e^+ e^-$} & CW &
        $1.37(7)$
        & &
        $1.11(6)$ 
        & &
        $1.13(6)$ 
        \\
        & VW &
        $1.36(7)$
        & &
        $1.17(6)$
        & & 
        $1.18(6)$ 
        \\
        \\[-0.25cm]
        \multirow{2}{*}{$\eta' \to \pi^0 \mu^+ \mu^-$} & CW &
        $0.92(5)$
        & &
        $0.610(35)$ 
        & &
        $0.603(35)$
        \\
        & VW &
        $0.90(5)$
        & &
        $0.64(4)$
        & &
        $0.65(4)$ 
        \\
        \midrule
        \multirow{2}{*}{$\eta' \to \eta e^+ e^-$} & CW &
        $4.77(7)$
        & &
        $4.38(6)$ 
        & &
        $4.41(6)$ 
        \\
        & VW &
        $4.65(7)$
        & &
        $4.56(7)$
        & &
        $4.56(7)$ 
        \\
        \\[-0.2cm]
        \multirow{2}{*}{$\eta' \to \eta \mu^+ \mu^-$} & CW &
        $2.60(6)$
        & &
        $1.93(4)$
        & &
        $1.88(4)$
        \\
        & VW &
        $2.54(5)$
        & &
        $2.05(4)$
        & &
        $2.18(4)$
        \\
		\bottomrule
	\end{tabular}
    }
	\caption{The same as \autoref{tab:BRs} but for the normalized branching ratios of the semileptonic decays, \autoref{eq:BR_normalized}.
    Due to partial cancellations in this ratio, the quoted uncertainties are given with the caveat that they are likely to underestimate the genuine uncertainty; see main text.}
	\label{tab:BRs_normalized}
\end{table}

\begin{figure*}[t]
	\centering
    \hspace{-1cm}
    \includegraphics[width=0.95\textwidth]{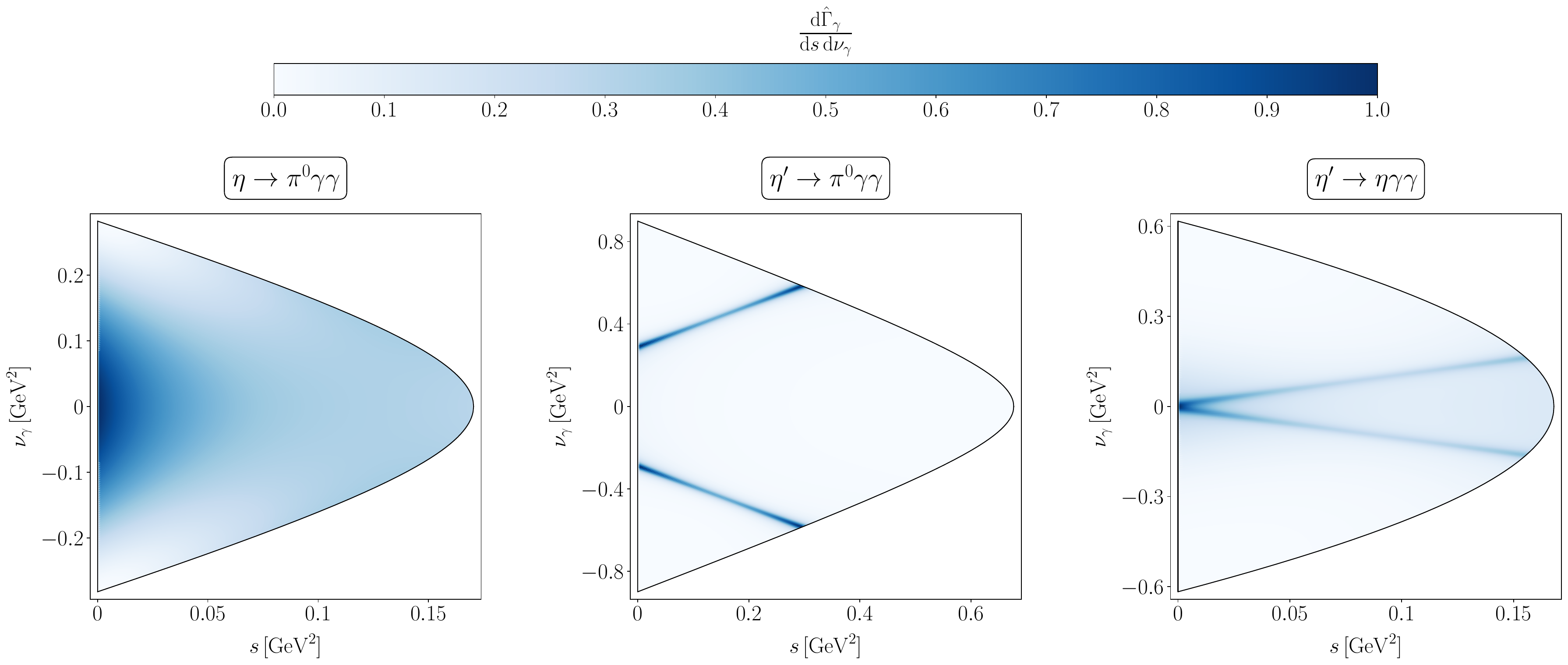}
    \\
    \includegraphics[width=0.325\textwidth]{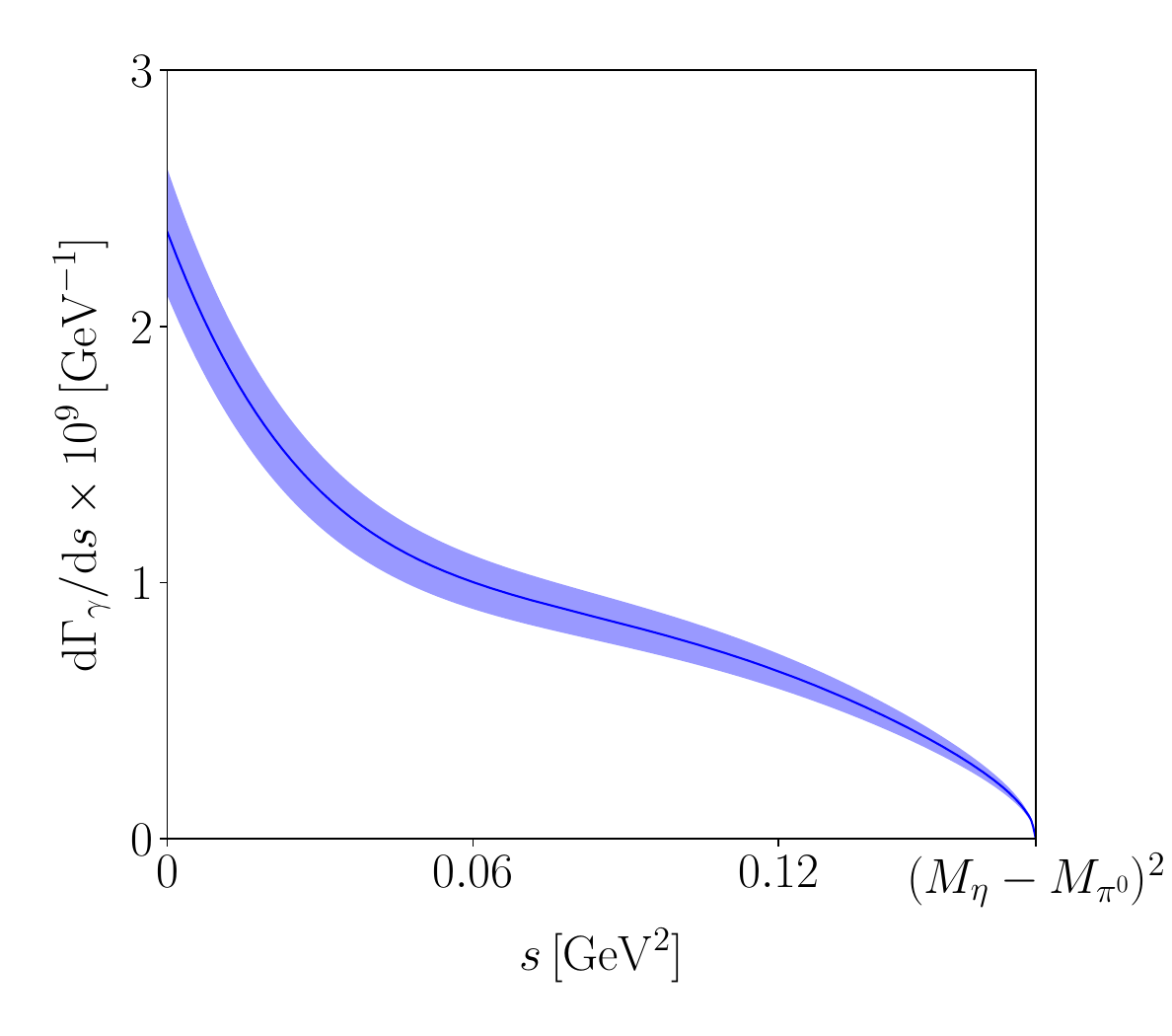}
    \includegraphics[width=0.325\textwidth]{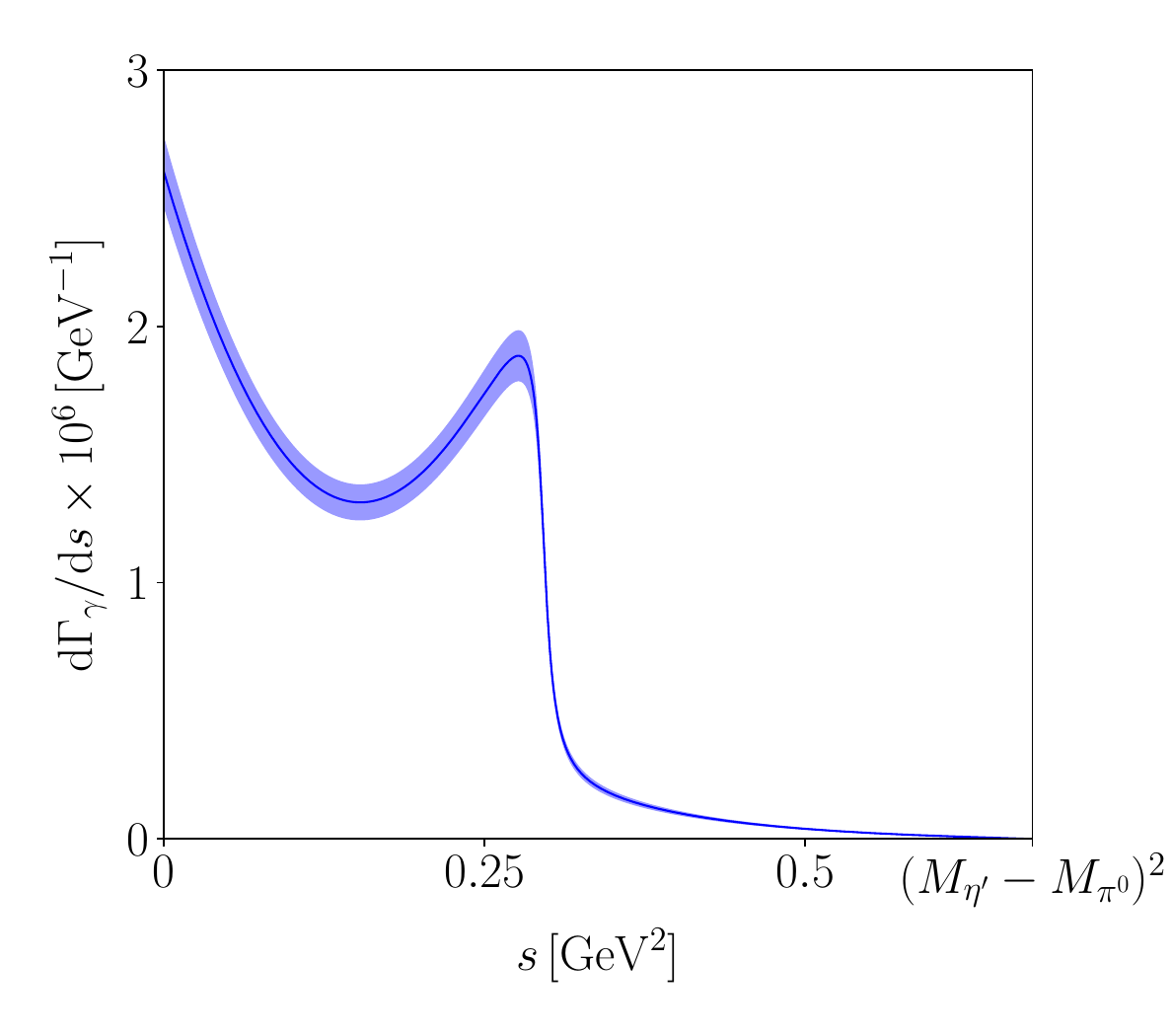}
    \includegraphics[width=0.325\textwidth]{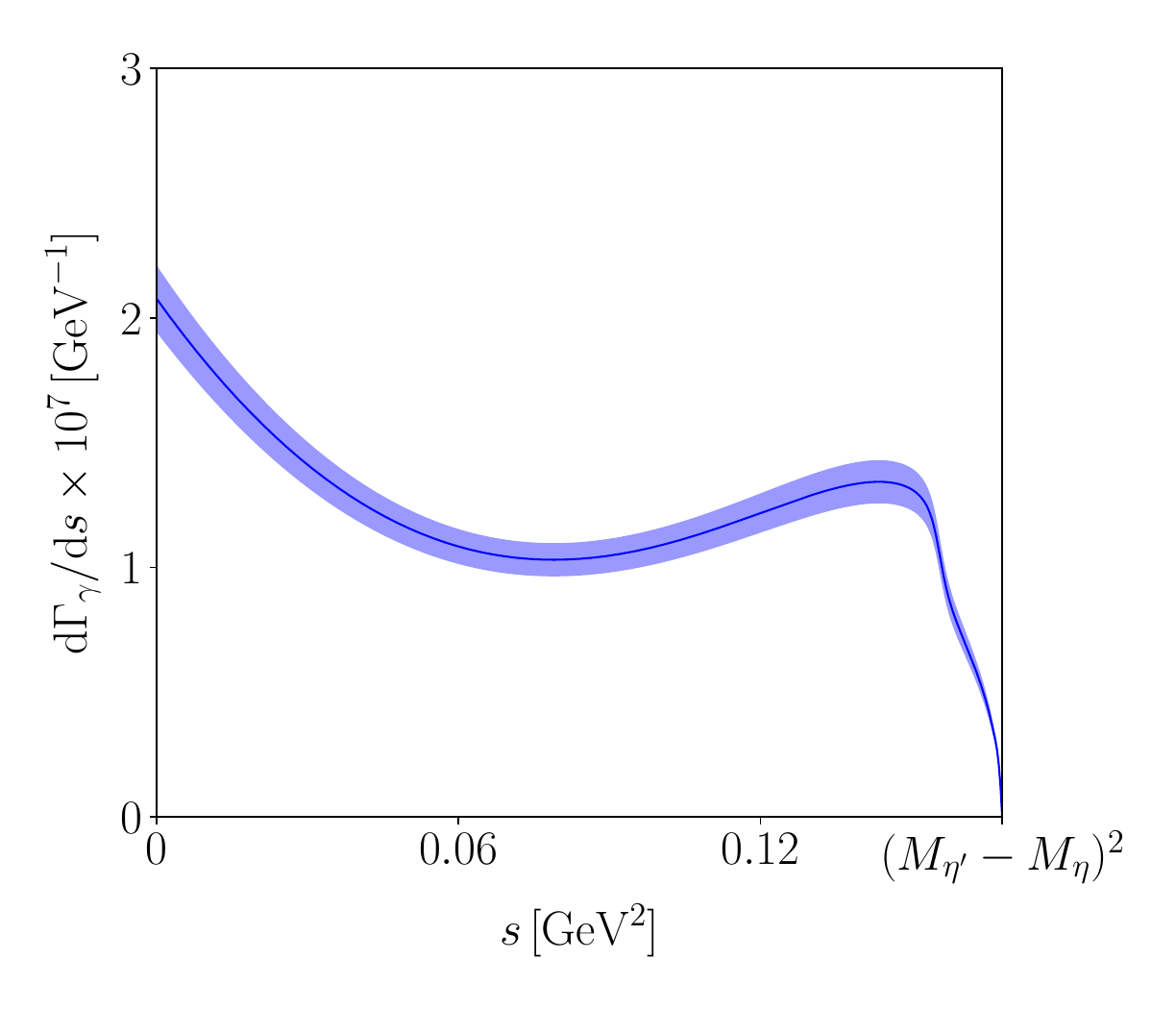}
	\caption{\LN{Dalitz} plots for the two-photon decays in the variant CW (\textit{top}), normalized to the maximum value within the available phase space of the respective channel, $\dff \hat{\Gamma}_\gamma/\dff s \, \dff \nu_\gamma = [\dff \Gamma_\gamma/\dff s \, \dff \nu_\gamma]/[\max\, \dff \Gamma_\gamma/\dff s \, \dff \nu_\gamma]$, and singly-differential decay widths in the \LN{Mandelstam} variable $s$, obtained in the variant CW (\textit{bottom}).}
	\label{fig:dalitz_diff_DW_gamma_gamma}
\end{figure*}
While our results for the two-photon decays of the $\eta'$ meson are compatible with the experimental results from BESIII,
$\BR(\eta' \to \pi^0 \gamma \gamma) = 3.20(24) \times 10^{-3}$~\cite{BESIII:2016oet} and 
$\BR(\eta' \to \eta \gamma \gamma) = 8.3(3.4)\times 10^{-5}$~\cite{BESIII:2019ofm},\footnote{%
    Here and in the following, we combine statistical and systematic uncertainties of experimental branching ratios in quadrature for simplicity.} 
the experimental situation for $\eta \to \pi^0 \gamma \gamma$ is presently inconclusive.
For this decay, the PDG average $\BR(\eta \to \pi^0 \gamma \gamma)=2.55(22) \times 10^{-4}$~\cite{ParticleDataGroup:2022pth}---the main input being $\BR(\eta \to \pi^0 \gamma \gamma) = 2.52(23) \times 10^{-4}$ from the A2 experiment at MAMI~\cite{A2atMAMI:2014zdf}---is in agreement with the theoretical calculation performed in Ref.~\cite{Lu:2020qeo} but in severe tension with the preliminary result from the KLOE-2 collaboration, $\BR(\eta \to \pi^0 \gamma \gamma) = 0.99(26) \times 10^{-4}$~\cite{Giovannella:2023}, which corroborates the older KLOE measurement $\BR(\eta \to \pi^0 \gamma \gamma) = 0.84(30) \times 10^{-4}$~\cite{KLOE:2005hln} and is consistent with the VMD-only result.
%$\BR(\eta' \to \pi^0 \gamma \gamma) = 3.20(7)(23) \times 10^{-3}$~\cite{BESIII:2016oet} % KEEP!
%$\BR(\eta' \to \eta \gamma \gamma) = 8.25(3.41)(0.72) \times 10^{-5}$~\cite{BESIII:2019ofm} % KEEP!
%$\BR(\eta \to \pi^0 \gamma \gamma) = 0.99(11)(24) \times 10^{-4}$~\cite{Giovannella:2023} % KEEP!
%$\BR(\eta \to \pi^0 \gamma \gamma) = 0.84(27)(14) \times 10^{-4}$~\cite{KLOE:2005hln} % KEEP!

The results for the normalized branching ratio can be found in \autoref{tab:BRs_normalized}, and the discussion of the differences between the distinct form-factor parameterizations is analogous to \autoref{sec:discussion_BRs}.
Due to partial cancellations in this ratio, the quoted uncertainties are reduced drastically, however with the caveat that they are likely to underestimate the genuine uncertainty, lest some neglected systematic effect beyond the error estimates of the couplings potentially becomes dominant here.
At the same time, potential corrections to the semileptonic branching ratios that are not included in the plain VMD model, \Lat{e.g.}, the $a_2$ resonance, are assumed to partially cancel as well because they emerge in the hadronic part of the amplitudes that is shared with the photonic decays.

The doubly- and singly-differential decay widths for the two-photon decays are displayed in \autoref{fig:dalitz_diff_DW_gamma_gamma}.
While the $\eta$ decay does not show much structure in either plot---being dominated by a $D$-wave at low and an $S$-wave at high diphoton invariant masses---the $\eta'$ decays are dominated by vector-meson resonances that can go quasi on-shell.
The $\omega$ resonance is clearly visible as two narrow bands in the \LN{Dalitz} plots and as a peak in the singly-differential distributions, whereas the $\rho$ is disguised in comparison due to its much larger width.
The angular dependence perceivable as a less saturated band in the \LN{Dalitz} plots and as a dip in the singly-differential distributions can be attributed to the fact that the $\omega \to [\pi^0/\eta] \gamma$ decay must be in a $P$-wave due to parity.

%---------------------------------------------------------------------------------------------------
\section{Summary}
\label{sec:summary}
%---------------------------------------------------------------------------------------------------
We have reanalyzed the standard-model contribution to the semileptonic decays $\eta^{(\prime)} \to \pi^0 \ell^+ \ell^-$ and $\eta' \to \eta \ell^+ \ell^-$, where $\ell=e, \mu$.
Since $\CC$ parity is conserved in the strong and electromagnetic interactions, these processes are mediated via a two-photon mechanism and therefore loop-induced.
This two-photon mechanism is known to be dominated by vector exchanges; as a major improvement compared to the existing literature, we have, for the first time, implemented a realistic dependence of the hadronic sub-process on the photon virtualities via vector-to-pseudoscalar transition form factors.
To assess the sensitivity to the chosen parameterizations, we compared three different schemes: constant couplings (as a reference point), monopole form factors, and dipole form factors.
The last of those three are motivated by having the correct asymptotic behavior at high virtualities.
In addition, dispersively improved variants of all form factors have been probed.
Non-trivial form factors turn out to be important in order not to overestimate the branching ratios.
We thereby improve previous theoretical results for the semileptonic $\eta^{(\prime)}$ decays.
On the other hand, the observables are mostly insensitive to the details of the parameterization at the level of uncertainty induced by the phenomenological coupling constants.

All predicted branching ratios are, as expected, well below the current experimental upper limits.
For the latter, we however recommend a reanalysis, given the far-from-flat \LN{Dalitz}-plot distributions of the standard-model contributions.
With improved experimental sensitivities in the future, our theoretical branching ratios of these rare $\eta^{(\prime)}$ decays can hopefully be compared to experiment and thus help cast a light on possible symmetry violations and physics beyond the standard model in the light-meson sector.

%---------------------------------------------------------------------------------------------------

\begin{acknowledgments}
We are grateful to Bernard \LN{Metsch} and Andreas \LN{Nogga} for valuable support regarding technical issues with \Lat{LoopTools} and \Lat{Collier}.
Furthermore, we thank Ansgar \LN{Denner} for insightful e-mail communication regarding \Lat{Collier}, and Martin \LN{Hoferichter}, Andrzej \LN{Kup\'s\'c}, Emilio \LN{Royo}, and Pablo \LN{S\'anchez-Puertas} for useful discussions.
Financial support by the DFG through the funds provided to the Sino--German Collaborative Research Center TRR110 ``Symmetries and the Emergence of Structure in QCD'' (DFG Project-ID 196253076 -- TRR 110) is gratefully acknowledged.
This work has been supported by STRONG-2020 ``The strong interaction at the frontier of knowledge: fundamental research and applications,'' which received funding from the European Union's Horizon 2020 research and innovation programme under grant agreement No.~824093.
Finally, the authors gratefully acknowledge the granted access to the bonna cluster hosted by the University of Bonn along with the support provided by its High Performance Computing \& Analytics Lab.
\end{acknowledgments}

%---------------------------------------------------------------------------------------------------

\appendix

%---------------------------------------------------------------------------------------------------
\section{\boldmath{$\Uthree$} flavor symmetry}
\label{appx:u3}
%---------------------------------------------------------------------------------------------------
For the $\Uthree$ parameterizations of the pseudoscalar and vector-meson multiplets, we write
\begin{align}\label{eq:multiplets}
	\Phi^P 
    &= 
    \begin{pmatrix}
        \pi^0 + \frac{\sqrt{2} \, \eta + \eta'}{\sqrt{3}} & 0 & 0 \\
        0 & \pi^0 + \frac{\sqrt{2} \, \eta + \eta'}{\sqrt{3}} & 0 \\
        0 & 0 & \frac{-\sqrt{2} \, \eta + 2 \eta'}{\sqrt{3}}
    \end{pmatrix},
    \notag \\
    \notag \\
	\Phi^{V^{(\prime)}}_\mu 
    &= 
    \begin{pmatrix}
        {\rho_\mu^0}^{(\prime)} + \omega_\mu^{(\prime)} & 0 & 0 \\
        0 & -{\rho_\mu^0}^{(\prime)} + \omega_\mu^{(\prime)} & 0 \\
        0 & 0 & -\sqrt{2} \phi_\mu^{(\prime)}
    \end{pmatrix}\hspace{-0.025cm},
\end{align}
where we only retain flavor-neutral states.
Here, mixing effects between the (physical) mesons are taken into account via the pattern
\begin{align}
	\begin{pmatrix}
		\eta' \\ \eta
	\end{pmatrix}
	&=
	\begin{pmatrix}
		\cos\theta_P & \sin\theta_P \\
		-\sin\theta_P & \cos\theta_P
	\end{pmatrix}
	\begin{pmatrix}
		\eta_1 \\ \eta_8
	\end{pmatrix},
	\notag \\
    \notag \\
	\begin{pmatrix}
		\omega^{(\prime)} \\ \phi^{(\prime)}
	\end{pmatrix}
	&=
	\begin{pmatrix}
		\cos\theta_{V^{(\prime)}} & \sin\theta_{V^{(\prime)}} \\
		-\sin\theta_{V^{(\prime)}} & \cos\theta_{V^{(\prime)}}
	\end{pmatrix}
	\begin{pmatrix}
		\omega_1^{(\prime)} \\ \omega_8^{(\prime)}
	\end{pmatrix},
\end{align}
with $\eta_1$, $\eta_8$ and $\omega_1^{(\prime)}$, $\omega_8^{(\prime)}$ denoting the isoscalar singlet and octet states of the pseudoscalar and vector-meson multiplets, respectively.
In the above, the mixing angles are assumed to be given by $\theta_P = \arcsin(-1/3)$ for the pseudoscalar nonet (canonical mixing) and $\theta_{V^{(\prime)}} = \arcsin(1/\sqrt{3})$ for the vector mesons (ideal mixing).
We furthermore introduce the charge matrix according to 
\beq
    \Q
    = \frac{1}{3} \diag[2,-1,-1].
\eeq

Using \autoref{eq:multiplets}, we calculate $\Tr[\Phi^P \Phi^V_\mu \Phi^{V^{(\prime)}}_\nu]$ to find the allowed couplings $\eta^{(\prime)} \rho \rho^{(\prime)}$, $\eta^{(\prime)} \omega \omega^{(\prime)}$, $\eta^{(\prime)} \phi \phi^{(\prime)}$, $\pi^0 \rho \omega^{(\prime)}$, and $\pi^0 \omega \rho^{(\prime)}$.
To derive the relative signs between the corresponding coupling constants $C_{V P \gamma}$ introduced in \autoref{sec:monopole}, we calculate $\Tr[\Phi^P \Phi^V_\mu \Q]$ and take the appropriate ratios of coefficients that emerge in \autoref{eq:amplitude}.
For our analysis, we furthermore included the OZI-suppressed coupling $C_{\phi \pi^0 \gamma}$, whose sign thus cannot be determined from $\Uthree$ symmetry.
Instead, we resort to analyses of $e^+ e^- \to 3\pi$ and $e^+ e^- \to \pi \gamma$~\cite{Hoferichter:2014vra,Hoferichter:2019mqg,Hoid:2020xjs}, which suggest that the product of the $\phi \gamma$ and $\phi \pi \gamma$ couplings carries a relative sign as compared to the product of the $\omega \gamma$ and $\omega \pi \gamma$ couplings. 
Hence, calculating $\Tr[\Phi^V_\mu \Q]$ indicates a relative sign between $C_{\phi \pi^0 \gamma}$ and $C_{\omega \pi^0 \gamma}$.
Fixing the sign of $C_{\rho \eta \gamma}$ to be positive, the sign convention of \autoref{tab:couplings_signs} follows.

\vfill \eject

%--------------------------------------------------------------------------------------------------
\section{Intermediate results}
\label{appx:intermediate_results}
%-------------------------------------------------------------------------------------------------
The numerical values of the auxiliary quantities $\Gamma_{V_1,V_2}^{(\gamma)}$ defined in \autoref{eq:DW_sum}, for a point-like interaction (PL), monopole form factors (MP), and dipole form factors (DP), are collected in \autoref{tab:DRs} and \autoref{tab:DRs_gamma_gamma}.
\begin{table*}[t]
	\centering
	\begin{tabular}{l  c  c  c  c  c  c  c  c}
	\toprule
		& & & $\Gamma_{\rho,\rho}/\MeV^5$ & $\Gamma_{\omega,\omega}/\MeV^5$ & $\Gamma_{\phi,\phi}/\MeV^5$ & $\Gamma_{\rho,\omega}/\MeV^5$ & $\Gamma_{\rho,\phi}/\MeV^5$ & $\Gamma_{\omega,\phi}/\MeV^5$ \\
		\midrule
        \multirow{6}{*}{$\eta \to \pi^0 e^+ e^-$} & \multirow{2}{*}{PL} & CW & $0.5302$ & \multirow{2}{*}{$0.5684$} & \multirow{2}{*}{$0.1864$} & $1.077$ & $0.6041$ & \multirow{2}{*}{$0.6485$} \\
        & & VW & $0.4992$ & & & $1.065$ & $0.6060$ & \\
        \\[-0.35cm]
  		& \multirow{2}{*}{MP} & CW & $0.3463$ & $0.3627$ & $0.1093$ & $0.6914$ & $0.3707$ & $0.3966$ \\
        & & VW & $0.3422$ & $0.3814$ & $0.1151$ & $0.7226$ & $0.3945$ & $0.4174$ \\
        \\[-0.35cm]
        & \multirow{2}{*}{DP} & CW & $0.3419$ & $0.3573$ & $0.1033$ & $0.6814$ & $0.3615$ & $0.3835$ \\
        & & VW & $0.3285$ & $0.3630$ & $0.09942$ & $0.7160$ & $0.3869$ & $0.3903$ \\
        \\[-0.15cm]
        \multirow{6}{*}{$\eta \to \pi^0 \mu^+ \mu^-$} & \multirow{2}{*}{PL} & CW & $0.3440$ & \multirow{2}{*}{$0.3686$} & \multirow{2}{*}{$0.1383$} & $0.7022$ & $0.4222$ & \multirow{2}{*}{$0.4498$} \\
        & & VW & $0.3123$ & & & $0.6785$ & $0.4136$ & \\
        \\[-0.35cm]
		& \multirow{2}{*}{MP} & CW & $0.1772$ & $0.1870$ & $0.06392$ & $0.3569$ & $0.2029$ & $0.2173$ \\
        & & VW & $0.1697$ & $0.1972$ & $0.06742$ & $0.3657$ & $0.2123$ & $0.2293$ \\
        \\[-0.35cm]
        & \multirow{2}{*}{DP} & CW & $0.1674$ & $0.1764$ & $0.05756$ & $0.3366$ & $0.1888$ & $0.2009$ \\
        & & VW & $0.1603$ & $0.1802$ & $0.06073$ & $0.3473$ & $0.1916$ & $0.2102$ \\
		\midrule
        \multirow{6}{*}{$\eta' \to \pi^0 e^+ e^-$} & \multirow{2}{*}{PL} & CW & $154.6$ & \multirow{2}{*}{$283.5$} & \multirow{2}{*}{$57.20$} & $405.1$ & $125.7$ & \multirow{2}{*}{$183.3$} \\ 
        & & VW & $152.8$ & & & $406.5$ & $138.9$ & \\
        \\[-0.35cm]
		& \multirow{2}{*}{MP} & CW & $125.8$ & $227.7$ & $37.08$ & $323.0$ & $82.41$ & $126.6$ \\ 
        & & VW & $133.7$ & $241.9$ & $39.93$ & $349.2$ & $103.0$ & $135.4$ \\
        \\[-0.35cm]
        & \multirow{2}{*}{DP} & CW & $128.1$ & $232.0$ & $35.95$ & $328.3$ & $84.42$ & $128.8$ \\ 
        & & VW & $131.5$ & $253.1$ & $38.66$ & $340.9$ & $101.0$ & $134.6$ \\
        \\[-0.15cm]
        \multirow{6}{*}{$\eta' \to \pi^0 \mu^+ \mu^-$} & \multirow{2}{*}{PL} & CW & $121.2$ & \multirow{2}{*}{$169.8$} & \multirow{2}{*}{$55.13$} & $284.5$ & $131.0$ & \multirow{2}{*}{$168.1$} \\ 
        & & VW & $116.9$ & & & $281.7$ & $139.1$ & \\
        \\[-0.35cm]
		& \multirow{2}{*}{MP} & CW & $80.02$ & $111.0$ & $30.42$ & $185.6$ & $70.21$ & $94.91$ \\ 
        & & VW & $83.84$ & $119.3$ & $32.79$ & $199.8$ & $84.77$ & $101.9$ \\
        \\[-0.35cm]
        & \multirow{2}{*}{DP} & CW & $79.10$ & $109.8$ & $28.68$ & $183.4$ & $68.78$ & $92.80$ \\ 
        & & VW & $80.95$ & $121.1$ & $29.78$ & $201.0$ & $82.23$ & $97.28$ \\
		\midrule
        \multirow{6}{*}{$\eta' \to \eta e^+ e^-$} & \multirow{2}{*}{PL} & CW & $19.68$ & \multirow{2}{*}{$50.07$} & \multirow{2}{*}{$6.701$} & $60.79$ & $8.303$ & \multirow{2}{*}{$14.86$} \\
        & & VW & $19.47$ & & & $60.64$ & $10.11$ & \\
        \\[-0.35cm]
		& \multirow{2}{*}{MP} & CW & $16.44$ & \multirow{2}{*}{$48.33$} & \multirow{2}{*}{$5.100$} & $48.56$ & $-1.684$ & \multirow{2}{*}{$10.79$} \\
        & & VW & $18.50$ & & & $57.98$ & $6.724$ & \\
        \\[-0.35cm]
        & \multirow{2}{*}{DP} & CW & $16.54$ & $51.24$ & $4.902$ & $47.02$ & $-2.518$ & $12.45$ \\
        & & VW & $18.37$ & $46.79$ & $4.827$ & $57.81$ & $6.109$ & $10.82$ \\
        \\[-0.15cm]
        \multirow{6}{*}{$\eta' \to \eta \mu^+ \mu^-$} & \multirow{2}{*}{PL} & CW & $12.45$ & \multirow{2}{*}{$20.56$} & \multirow{2}{*}{$4.847$} & $31.57$ & $10.52$ & \multirow{2}{*}{$15.70$} \\
        & & VW & $12.38$ & & & $31.86$ & $11.66$ & \\
        \\[-0.35cm]
        & \multirow{2}{*}{MP} & CW & $8.240$ & \multirow{2}{*}{$16.03$} & \multirow{2}{*}{$3.170$} & $19.66$ & $2.342$ & \multirow{2}{*}{$9.959$} \\
        & & VW & $9.471$ & & & $24.59$ & $6.988$ & \\
        \\[-0.35cm]
        & \multirow{2}{*}{DP} & CW & $7.980$ & $16.28$ & $2.944$ & $18.15$ & $1.682$ & $10.13$ \\
        & & VW & $10.05$ & $15.35$ & $2.937$ & $23.61$ & $6.266$ & $9.555$ \\
		\bottomrule
	\end{tabular}
    \caption{Numerical results for the auxiliary quantities defined in \autoref{eq:DW_sum} for the models PL, MP, and DP in both variants CW and VW, rounded to four significant digits.}
	\label{tab:DRs}
\end{table*}
\begin{table*}[t]
	\centering
	\begin{tabular}{l  c  c  c  c  c  c  c}
	\toprule
		& & $\Gamma_{\rho,\rho}^\gamma/\MeV^5$ & $\Gamma_{\omega,\omega}^\gamma/\MeV^5$ & $\Gamma_{\phi,\phi}^\gamma/\MeV^5$ & $\Gamma_{\rho,\omega}^\gamma/\MeV^5$ & $\Gamma_{\rho,\phi}^\gamma/\MeV^5$ & $\Gamma_{\omega,\phi}^\gamma/\MeV^5$ \\
		\midrule
		\multirow{2}{*}{$\eta \to \pi^0 \gamma \gamma$} & CW & $3.154 \times 10^4$ & \multirow{2}{*}{$3.193 \times 10^4$} & \multirow{2}{*}{$8.719 \times 10^3$} & $6.175 \times 10^4$ & $3.218 \times 10^4$ & \multirow{2}{*}{$3.335 \times 10^4$} \\
        & VW & $2.921 \times 10^4$ & & & $6.108 \times 10^4$ & $3.189 \times 10^4$ & \\
		\midrule
		\multirow{2}{*}{$\eta' \to \pi^0 \gamma \gamma$} & CW & $3.088 \times 10^7$ & \multirow{2}{*}{$4.586 \times 10^8$} & \multirow{2}{*}{$4.286 \times 10^6$} & $1.097 \times 10^8$ & $1.115 \times 10^7$ & \multirow{2}{*}{$1.884 \times 10^7$} \\ 
        & VW & $3.341 \times 10^7$ & & & $1.130 \times 10^8$ & $1.386 \times 10^7$ & \\
		\midrule
		\multirow{2}{*}{$\eta' \to \eta \gamma \gamma$} & CW & $3.203 \times 10^6$ & \multirow{2}{*}{$6.537 \times 10^7$} & \multirow{2}{*}{$4.473 \times 10^5$} & $1.425 \times 10^7$ & $2.031 \times 10^5$ & \multirow{2}{*}{$7.406 \times 10^5$} \\ 
        & VW & $3.280 \times 10^6$ & & & $1.411 \times 10^7$ & $5.056 \times 10^5$ & \\
		\bottomrule
	\end{tabular}
    \caption{Numerical results for the auxiliary quantities defined in \autoref{eq:DW_sum} in both variants CW and VW, rounded to four significant digits.}
	\label{tab:DRs_gamma_gamma}
\end{table*}

%---------------------------------------------------------------------------------------------------
\section{Constants and parameters}
\label{appx:constants}
%---------------------------------------------------------------------------------------------------
We collect the masses and widths used throughout the calculations in \autoref{tab:constants}.
\begin{table}[t]
	\centering
	\begin{tabular}{c  c  r}
	\toprule
		 & Variable & Value~\cite{ParticleDataGroup:2022pth} \\
		\midrule
		$\pi^0$ & $\MpiN$ & $134.9768(5) \MeV$ \\
		$\pi^\pm$ & $\MpiC$ & $139.57039(18) \MeV$ \\
		$K^0$ & $\MK$ & $497.611(13) \MeV$ \\
		\multirow{2}{*}{$\eta$} & $\Meta$ & $547.862(17) \MeV$ \\ 
        & $\Gamma_\eta$ & $1.31(5)\keV$ \\
		\multirow{2}{*}{$\eta'(958)$} & $\MetaPrime$ & $957.78(6) \MeV$ \\
		& $\GetaPrime$ & $188(6) \keV$ \\
		\midrule
		\multirow{2}{*}{$\rho^0(770)$} & $\Mrho$ & $775.26(23) \MeV$ \\
		& $\Grho$ & $147.4(8) \MeV$ \\
		\multirow{2}{*}{$\omega(782)$} & $\Momega$ & $782.66(13) \MeV$ \\
		& $\Gomega$ & $8.68(13) \MeV$ \\
        ${K^*}^0(892)$ & $\MKStar$ & $895.55(20) \MeV$ \\
		\multirow{2}{*}{$\phi(1020)$} & $\Mphi$ & $1019.461(16) \MeV$ \\
		& $\Gphi$ & $4.249(13) \MeV$ \\
		\multirow{2}{*}{$\rho^0(1450)$} & $\MrhoPrime$ & $1465(25) \MeV$ \\
		& $\GrhoPrime$ & $400(60) \MeV$ \\
		\multirow{2}{*}{$\omega(1420)$} & $\MomegaPrime$ & $1410(60) \MeV$ \\
		& $\GomegaPrime$ & $290(190) \MeV$ \\
		\multirow{2}{*}{$\phi(1680)$} & $\MphiPrime$ & $1680(20) \MeV$ \\
		& $\GphiPrime$ & $150(50) \MeV$ \\
		\bottomrule
	\end{tabular}
	\caption{The masses and widths needed for the calculations in this article, with the values taken from Ref.~\cite{ParticleDataGroup:2022pth}.
	}
	\label{tab:constants}
\end{table}

%---------------------------------------------------------------------------------------------------
% Bib
%---------------------------------------------------------------------------------------------------

\bibliographystyle{apsrev4-2_mod}

\bibliography{Bib/bibliography} 

\end{document}